\documentclass[twocolumn]{aastex63}
\usepackage{multirow}
\usepackage[figuresright]{rotating}
\usepackage{subfigure}
\usepackage{epic,eepic}
\usepackage{graphicx}
\usepackage{pifont}

\shorttitle{}
\shortauthors{}

\begin{document}

\title{Discovery of one neutron star candidate from radial velocity monitoring}

\correspondingauthor{Song Wang}
\email{songw@bao.ac.cn}

\author{Hailong Yuan}
\affiliation{Key Laboratory of Optical Astronomy, National Astronomical Observatories, Chinese Academy of Sciences, Beijing 100101, China}
\affiliation{College of Astronomy and Space Sciences, University of Chinese Academy of Sciences, Beijing 100049, China}

\author{Song Wang}
\affiliation{Key Laboratory of Optical Astronomy, National Astronomical Observatories, Chinese Academy of Sciences, Beijing 100101, China}
\affiliation{College of Astronomy and Space Sciences, University of Chinese Academy of Sciences, Beijing 100049, China}

\author{Zhongrui Bai}
\affiliation{Key Laboratory of Optical Astronomy, National Astronomical Observatories, Chinese Academy of Sciences, Beijing 100101, China}
\affiliation{College of Astronomy and Space Sciences, University of Chinese Academy of Sciences, Beijing 100049, China}

\author{Yue Wang}
\affiliation{Key Laboratory of Optical Astronomy, National Astronomical Observatories, Chinese Academy of Sciences, Beijing 100101, China}
\affiliation{College of Astronomy and Space Sciences, University of Chinese Academy of Sciences, Beijing 100049, China}

\author{Yiqiao Dong}
\affiliation{Key Laboratory of Optical Astronomy, National Astronomical Observatories, Chinese Academy of Sciences, Beijing 100101, China}

\author{Mengxin Wang}
\affiliation{Key Laboratory of Optical Astronomy, National Astronomical Observatories, Chinese Academy of Sciences, Beijing 100101, China}

\author{Sicheng Yu}
\affiliation{Key Laboratory of Optical Astronomy, National Astronomical Observatories, Chinese Academy of Sciences, Beijing 100101, China}

\author{Yongheng Zhao}  
\affiliation{Key Laboratory of Optical Astronomy, National Astronomical Observatories, Chinese Academy of Sciences, Beijing 100101, China}
\affiliation{College of Astronomy and Space Sciences, University of Chinese Academy of Sciences, Beijing 100049, China}

\author{Yaoquan Chu}
\affiliation{University of Science and Technology of China, Hefei 230026, China}

\author{Jifeng Liu}
\affiliation{Key Laboratory of Optical Astronomy, National Astronomical Observatories, Chinese Academy of Sciences, Beijing 100101, China}
\affiliation{College of Astronomy and Space Sciences, University of Chinese Academy of Sciences, Beijing 100049, China}
\affiliation{WHU-NAOC Joint Center for Astronomy, Wuhan University, Wuhan, Hubei 430072, China}

\author{Haotong Zhang}
\affiliation{Key Laboratory of Optical Astronomy, National Astronomical Observatories, Chinese Academy of Sciences, Beijing 100101, China}
\affiliation{College of Astronomy and Space Sciences, University of Chinese Academy of Sciences, Beijing 100049, China}

\begin{abstract}

We report the discovery of one possible neutron star binary ($P_{\rm orb} =$ 0.8666 day) by using the LAMOST low-resolution spectroscopic data.
The visible companion is a late A-type dwarf 
($T_{\rm eff} = 7900 \pm 200$ K; log$g$ $=$ 4.3$\pm$0.2; $M =$ 1.7$\pm$0.1 M$_{\odot}$; $R\ =\ 1.7\pm0.2$ R$_{\odot}$),  at a distance of 1.11$\pm0.03$ kpc.
No double-lined feature can be seen from the GTC/HORuS high-resolution spectra, thus the radial velocity variation indicates an invisible object hiding in the binary.
The system's optical light curves show clear ellipsoidal variability, suggesting that the visible companion is tidal distorted.
By fitting the multi-band light curves with the ELC and WD codes, we constrain the mass of the invisible star to be 1.1--1.3 M$_{\odot}$.
Spectral disentangling shows no additional component with optical absorption spectra, supporting the system contains one compact object.
No X-ray or UV emission are detected in the ROSAT archive observations.
Therefore, we suspect the invisible object is more likely a neutron star rather than a white dwarf.
Our finding suggests the ability of LAMOST spectroscopic survey to discover X-ray quiescent compact objects.

\end{abstract}


\section{INTRODUCTION}
\label{intro.sec}

Radial velocity (RV) monitoring has been proved to be one feasible way to discover compact objects in binary systems.
Spectral observations of the normal star companion provide a direct measurement of the dynamical mass of the unseen object.
Theory predicts that the majority of binaries containing one normal star and one compact object do not emit X-ray. Recently, about 10 X-ray quiescent star-black hole (BH) were found through RV measurement of the companion star \citep{2014Natur.505..378C,2019Sci...366..637T,2019Natur.575..618L,2020A&A...637L...3R,2021arXiv210102212J}. Although some candidates are sill under debate, the number of which has reached half of those stellar-mass BHs discovered from X-rays  over about fifty years \citep{2006ARA&A..44...49R}.
More star-neutron star (NS) binaries are expected to exist than star-BH systems, although the discovery may be more difficult due to smaller RV variation.
Recently, several possible dormant NSs have been discovered by using the RV monitoring method \citep[e.g.,][]{2022arXiv220611270M, 2022NatAs.tmp..201Y}, evidencing the capability of discovering non-accreting NSs in binaries from optical time-domain (TD) spectroscopic survey.

NSs evolved through different evolutionary paths have different mass distribution \citep[e.g., peak and cutoff values;][]{2013ApJ...778...66K}.
The NSs in double NS and NS-white dwarf systems show peaks at 1.33 and 1.55 M$_{\odot}$, respectively, with the difference (0.22 M$_{\odot}$) suggesting a mass accretion occurred during the spin-up phase.
The mass cutoff are at $\approx$1.6 M$_{\odot}$ for NSs in double NS systems, while at $\approx$2.1 M$_{\odot}$ for NSs with white dwarf companions.
The most massive NSs known have masses ranging from 2 to 2.4
\citep[e.g.,][]{2011ApJ...728...95V,2013Sci...340..448A,2018ApJ...859...54L,2020NatAs...4...72C}, and their masses were determined through RV measurement or relativistic Shapiro delay.
Some objects show higher masses around 3 M$_{\odot}$, and their nature (either the most massive NS or the least massive BH) are ambiguous.
\citet{2019Sci...366..637T} reported a massive (3.3$^{+2.8}_{-0.7}$ M$_{\odot}$) unseen companion in a noninteracting binary, which may be a low-mass BH or a massive NS.
The nature of the remnant of GW170817 \citep{2017PhRvL.119p1101A}, from a merger of two NSs, is still under debate \citep{2017ApJ...850L..19M,2018ApJ...861..114Y,2018ApJ...859L..23P,2019PhRvD.100b3015S}.
Therefore, the discovery of more NSs and stellar-mass BHs can shed light on their mass distribution, which are greatly helpful in understanding the evolution of massive stars.

Recently, by using the LAMOST (Large Sky Area Multi-Object fiber Spectroscopic Telescope; also known as GuoShouJing telescope) spectroscopic survey data, we identify one star (2MASS J06163552+2319094, hereafter J06163552; TIC 429716088) showing periodic RV variation.
The coordinate is (R.A., Decl.) = (06:16:35.52, 23:19:09.3), with $V$-band magnitude of 13.23 mag.
The ASAS-SN\footnote{https://asas-sn.osu.edu/variables} and ZTF\footnote{https://www.ztf.caltech.edu/} light curves (LCs) shows a period of 0.8666 day, with an ellipsoidal shape (Figure \ref{show.fig}).
By combining the photometric and spectroscopic data analysis, we confirm that J06163552 is a NS. In Section \ref{obs.sec}, we present the photometric and spectroscopic observation and data reduction. Section \ref{star.sec} shows the information of the visible star, including stellar parameters, distance, and masses, etc. In Section \ref{orbit.sec}, we perform RV fitting using {\it The Joker} code and LC fitting using the ELC code and WD code. Finally, we give a discussion and summary in Section \ref{diss.sec} and \ref{sum.sec}.

\section{Observations and data reduction}
\label{obs.sec}

\subsection{Photometry}

To obtain multi-band LCs of J06163552 with higher accuracy than ASAS-SN and ZTF data, we carried out imaging observations on five nights (Jan. 5/6/7, 2020; Feb. 5/7, 2020) with the 85-cm telescope at the Xinglong station of the National Astronomical Observatories of China (NAOC). It is equipped with a 1024$\times$1024 pixel CCD and Johnson-Cousin-Bessell $UBVR_{\rm c}I_{\rm c}$ filter system \citep{2009RAA.....9..349Z}.
We made bias and flat-field corrections to the raw images. Then we use the Python module {\sf Photutils} to do aperture photometry. Differential photometry was performed with nearby bright non-variable stars.
Absolute magnitudes were derived by correcting the median magnitudes to literature catalogues: $B$ and $V$ magnitudes from the UCAC4 catalogue \citep{2012yCat.1322....0Z}, 
and $R$ and $I$ magnitude from the USNO-B1.0 catalogue \citep{2003AJ....125..984M}.

\begin{figure}[htbp!]
\center
\includegraphics[width=0.48\textwidth]{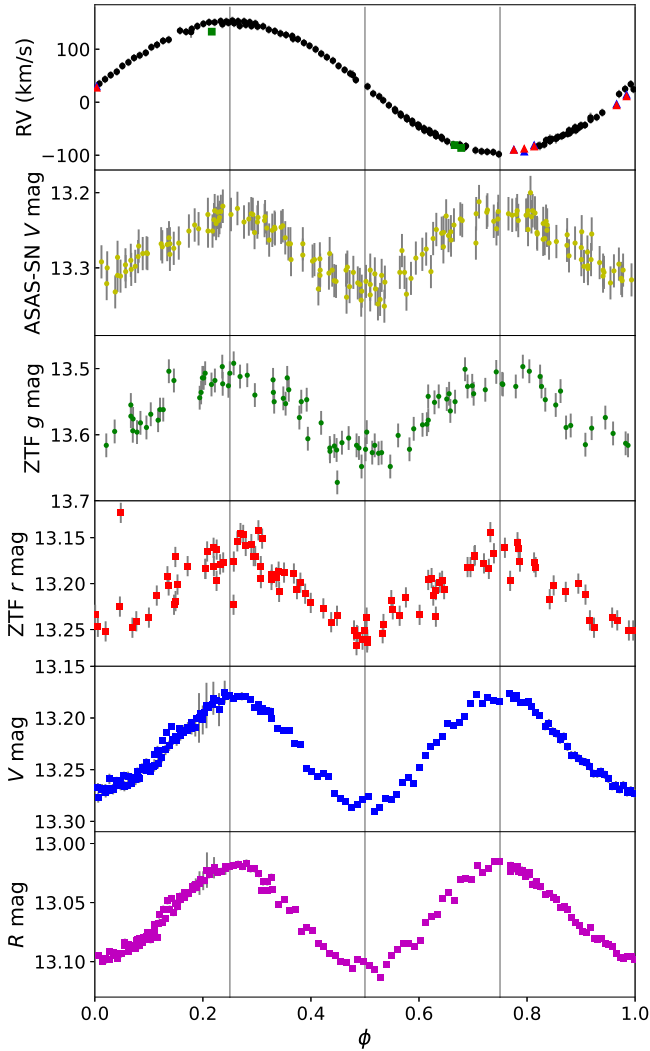}
\caption{Folded RV curves and LCs in different bands ($V$ band of ASAS-SN, $g$ and $r$ bands of ZTF, $V$ and $R$ bands of the Xinglong 85-cm telescope), with a period of 0.866 day. The phases of the inferior conjunction ($\phi =$ 0 or 1), quadrature ($\phi =$ 0.25 and 0.75), and superior conjunction  $\phi =$ 0.5 are marked with vertical gray lines. The RV data are determined from the LAMOST low-resolution spectra (black circles), the LAMOST med-resolution $B$- and $R$-band spectra (blue and red triangles, respectively), and the HORuS high-resolution spectra (green squares).}
\label{show.fig}
\end{figure}

For clarity we named the visible star, which shows periodic RV and LC modulation, as star 1 or the primary, and marked the unseen component as star 2 or the secondary.
The ephemeris of the system is
\begin{equation}
T (\phi = 0) = 2457690.3922(5)BJD + 0.8666(1) \times N,
\end{equation}
where $\phi =$  0 corresponds to the primary star in inferior conjunction with the secondary object, and $BJD$ is the Barycentric Julian Date.
The folded LCs show the characteristic double-peaked morphology expected for a tidally distorted primary (Figure \ref{show.fig}), with two equally bright peaks when the system is viewed at quadrature ($\phi =$ 0.25 and 0.75). Due to gravity darkening, the primary is fainter at $\phi =$ 0.5 (at its superior conjunction) than at $\phi =$ 0 or 1 (at its inferior conjunction).

\subsection{Spectroscopy}

\subsubsection{LAMOST observation}

LAMOST is a reflecting Schmidt telescope, with an effective aperture of 4 m and a field of view of 5 degrees
\citep{2012RAA....12.1197C,2012RAA....12..723Z}.
From 2011 to 2021, LAMOST\footnote{http://www.lamost.org/} has released over 10 million low-resolution (R $\approx$ 1800) spectra and over 14 million med-resolution (R $\approx$ 7500) spectra \citep[][]{2022Innov...300224Y}.
Each public low-resolution spectrum is a coadded spectrum combined by three single exposures.
Besides, LAMOST performed TD spectroscopic observations for some plates \citep[e.g.][]{2021RAA....21..292W}, each of which have numerous exposures. 
{\bf T}he plate containing J06163552 gained 210 low-resolution exposures from Mar. 2015 to Jan. 2019, and gained 13 med-resolution exposures from Dec. 2020 to Feb. 2021. This plate, containing about 3000 targets, was chosen from the $Kepler$ $K$2 Campaign 0 since it includes a lot of photometric variables or RV variables. One of these targets, LB-1, has been proposed to be a quiescent BH \citep{2019Natur.575..618L}.
The data for the whole plate will be released soon (Bai et al in preparation).
The wavelength of the low-resolution spectrum (LRS) ranges from 3,690 \AA\ to 9,100 \AA\ 
\citep{2015RAA....15.1095L}, while for the med-resolution spectrum (MRS), the blue and red arms cover wavelength ranges from 4950 \AA\ to 5350 \AA\ and from 6300 \AA\ to 6800 \AA, respectively \citep{2020arXiv200507210L}.

The raw CCD data were reduced by the LAMOST 2D pipeline, including bias and dark subtraction, flat field correction, spectrum extraction, sky background subtraction, wavelength calibration, etc \citep[see][for details]{2017PASP....17.91}.
The wavelength calibration of the data was based on the Sr and ThAr lamps and night sky lines \citep{2010ApJ...718.1378M}.
The released spectra used the vacuum wavelength scale and had been corrected to the heliocentric frame.

In order to derive accurate RV values, the spectra with low signal-to-noise ratio ($SNR$) ($g$-band $SNR<$ 10 or $r$-band $SNR<$ 10) were removed.
For LRS, we measured RV with the classical cross-correlation technique,
i.e. by shifting and comparing the best matched template to the observed spectrum.
The wavelength bands  to calculate the cross-correlation function (CCF) are 4000--5300\AA\ and 6350--7000\AA.
The Kurucz model \citep[][]{Castelli2003} with the parameter of $T_{\mathrm{eff}}=7600$ K, $\mathrm{log}\textit{g}=4.0$ and $[\mathrm{Fe/H}]=0.0$ was used as the template.
First, the RV grid added to the template is $-$1500 km/s $\sim$ 1500 km/s with a step of $\Delta{\log\lambda}=0.0001$ ($\Delta{v}=69$ km/s).
The RV with minimum $\chi^2$ ($\nu_b$) was selected.
Second, in the range of $v_b-138$ km/s $\sim$ $v_b+138$ km/s, we used a grid of $\Delta{\log\lambda}=0.00002$ ($\Delta{v}=14$ km/s).
The bottom of the $\chi^2$ curve was fitted with a Gaussian function to find the minimum value (corresponding to the final RV) and the fitting uncertainty.
The measurement error of the RV is the quadratic sum of the $\sigma$ of the Gaussian function and the formal error where $\Delta\chi^2=1$.
For MRS, we measured the RV values by using the CCF method with the $B$- and $R$-band spectra, respectively.
The RV uncertainty is the quadratic sum of the wavelength calibration uncertainty and the measurement error.
For an A-type star, the wavelength calibration uncertainty of the LRS with SNR above 40 is about 7.28 km/s \citep{2021RAA....21..249B}; while for the MRS, the uncertainty is around 1 km/s.
The RV measurements are listed in Table \ref{lamost.tab} and \ref{lamostm.tab}.

\subsubsection{HORuS observation}
\label{subsect:horus}

The High Optical Resolution Spectrograph (HORuS), a moderate resolution echelle spectrograph ($R$ $\approx$ 25,000), is in operation at the 10.4 m Gran Telescopio Canarias (GTC). The wavelength coverage ranges from 370 nm to 690 nm.

\begin{figure*}[htbp!]
\center
\includegraphics[width=1\textwidth]{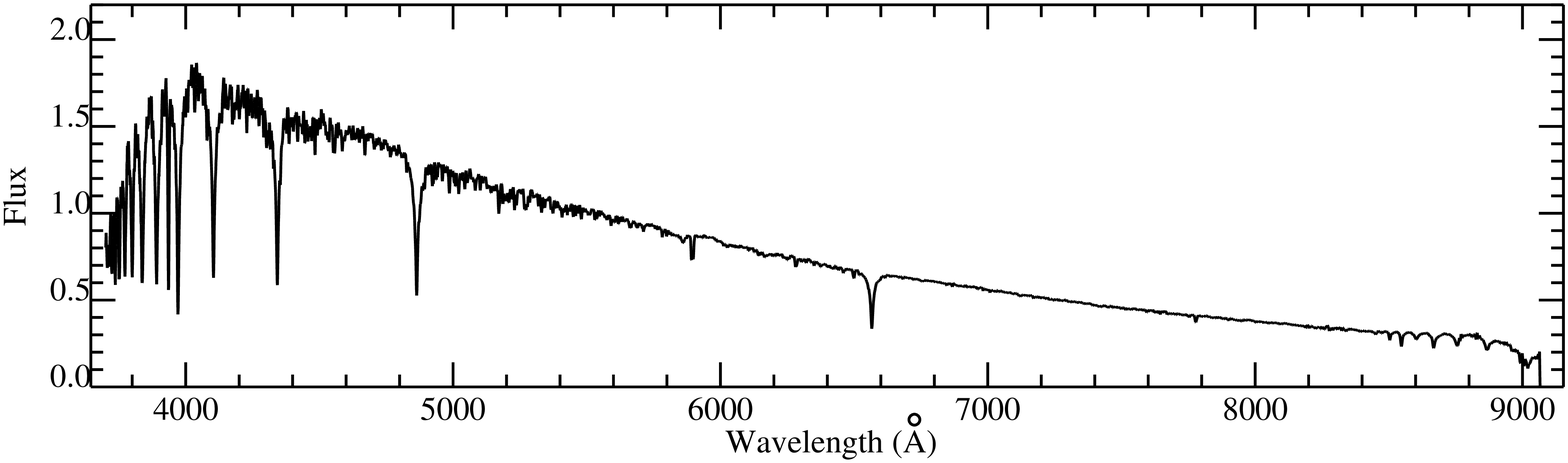}
\includegraphics[width=1\textwidth]{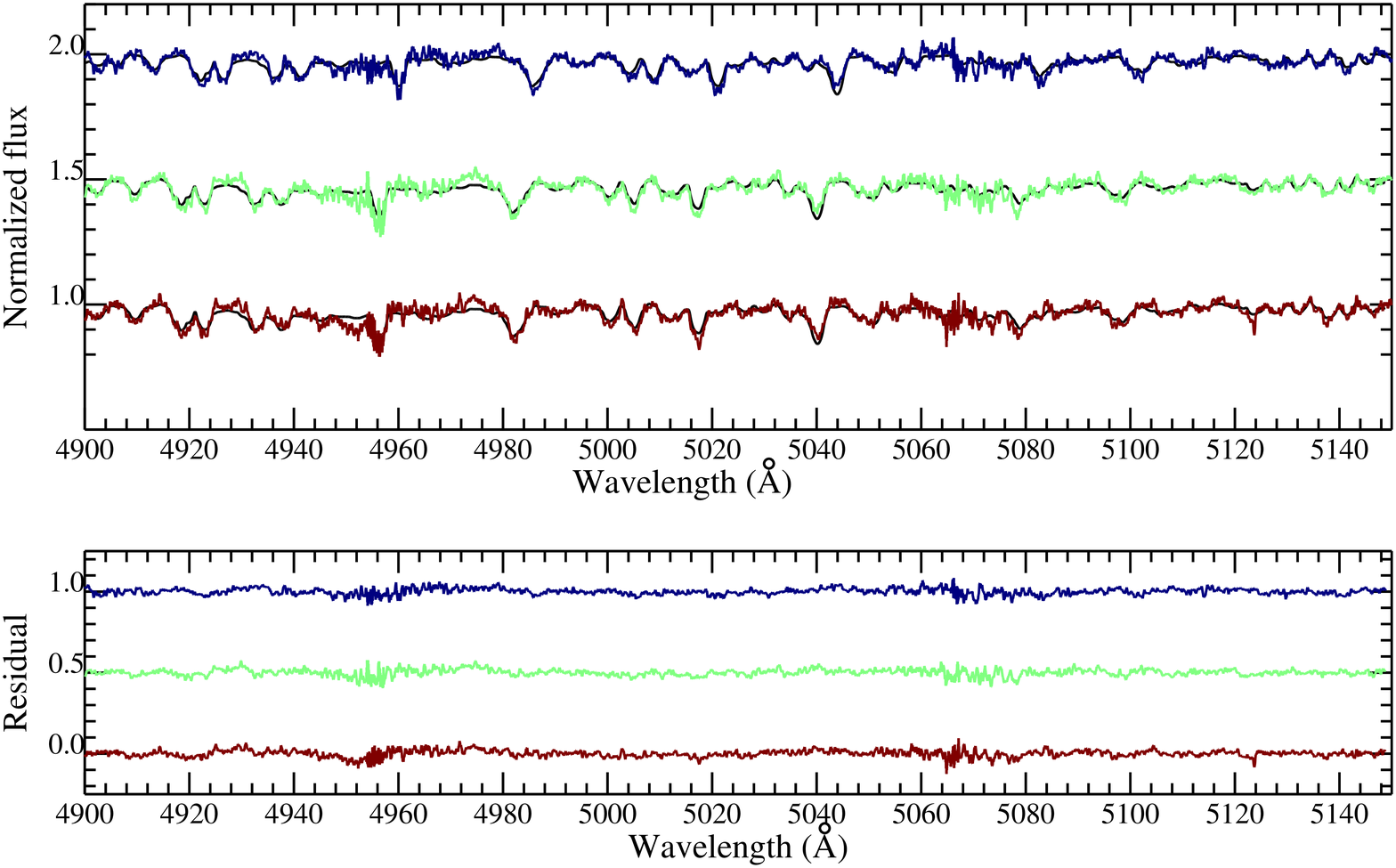}
\caption{
Upper panel: One low resolution spectrum of J06163552 taken by LAMOST on Nov 7th, 2016.
Middle panel: The three HORuS high-resolution spectra, of which the flux was shifted for convenience. The spectra are in chronological order from bottom to top (see Table \ref{horus.tab}), and they were observed at the phase $\phi \approx$ 0.67, 0.68, and 0.22, respectively. The doppler-shifted model spectra ($T_{\rm eff}=7900$ K, log$g=4.2$ and [Fe/H] $=$ 0) are marked with black lines.
Lower panel: Residuals showing the difference between the observed spectra and the template. No double-line feature can be seen.}
\label{spfig}
\end{figure*}

J06163552 was observed three times (Nov. 21/26, 2019) with HORuS.
We used the {\bf chain}\footnote{https://github.com/callendeprieto/chain}, a fully automated pipeline written in IDL for reducing echelle spectra, to process the HORuS data.
It performs bias removal, cosmic ray cleaning, order tracing, extraction, and wavelength calibration using the Th-Ar lamp.
The RV measurement process is the same as above, and the results are listed in Table \ref{horus.tab}.
The spectra of HORuS shows J06163552 is s single-line spectroscopic binary (Figure \ref{spfig}).

\begin{table}
\caption{Barycentric-corrected RV values from LAMOST low-resolution observations. \label{lamost.tab}}
\setlength{\tabcolsep}{4.5pt}
\begin{center}
 \begin{tabular}{cc}
\hline\noalign{\smallskip}
BMJD & RV \\
(day) & (km/s) \\
\hline\noalign{\smallskip}
57085.54305 & $-53.44\pm7.53$\\
57085.55277 & $-58.76\pm7.57$\\
57438.55458 & $15.42\pm7.88$\\
57438.56430 & $25.06\pm7.73$\\
57438.57333 & $34.02\pm7.87$\\
57699.74350 & $124.73\pm7.36$\\
57699.75253 & $118.92\pm7.37$\\
57699.76226 & $112.04\pm7.35$\\
57699.77128 & $106.45\pm7.36$\\
57699.78101 & $99.80\pm7.37$\\
57699.79004 & $91.57\pm7.38$\\
57699.79906 & $84.92\pm7.38$\\
57699.80879 & $78.40\pm7.40$\\
57699.81782 & $68.50\pm7.38$\\
57699.82754 & $60.94\pm7.37$\\
57699.83657 & $54.11\pm7.35$\\
57699.84560 & $41.99\pm7.41$\\
57699.86435 & $29.71\pm7.37$\\
57699.87338 & $16.27\pm7.41$\\
57699.88310 & $10.89\pm7.41$\\
\noalign{\smallskip}\hline
\end{tabular}
\end{center}
\smallskip
{This table is available in its entirety in machine-readable and Virtual Observatory (VO) forms in the online journal. A portion is shown here for guidance regarding its form and content.}
\end{table}

\begin{table}
\caption{Barycentric-corrected ($B$- and $R$-band) RV values from LAMOST med-resolution observations. \label{lamostm.tab}}
\setlength{\tabcolsep}{4.5pt}
\begin{center}
 \begin{tabular}{ccc}
\hline\noalign{\smallskip}
BMJD & RV$_b$ & RV$_l$ \\
(day) & (km/s) & (km/s) \\
\hline\noalign{\smallskip}
59236.58082 & $-89.45\pm1.04$ & $-89.97\pm1.31$\\
59236.59749 & $-93.23\pm4.26$ & $-87.79\pm1.78$\\
59236.61346 & $-81.03\pm1.01$ & $-83.63\pm1.68$\\
59264.47686 & $-3.47\pm1.79$ & $-5.89\pm1.87$\\
59264.49283 & $14.07\pm1.98$ & $11.01\pm2.32$\\
59264.50950 & $29.94\pm1.01$ & $27.29\pm1.35$\\
\noalign{\smallskip}\hline
\end{tabular}
\end{center}
\end{table}

\begin{table}
\caption{Barycentric-corrected RV values from HORuS observations. \label{horus.tab}}
\setlength{\tabcolsep}{4.5pt}
\begin{center}
 \begin{tabular}{cc}
\hline\noalign{\smallskip}
BMJD & RV  \\
(day) & (km/s)  \\
\hline\noalign{\smallskip}
58809.25208 & $-80.91\pm$4.36\\
58809.26249 & $-86.31\pm$4.59\\
58814.06205 & $133.12\pm$4.46\\
\noalign{\smallskip}\hline
\end{tabular}
\end{center}
\end{table}

\section{The visible star}
\label{star.sec}

\subsection{Stellar parameters}
\label{para.sec}

LAMOST DR8 gives atmospheric parameters of the effective temperature $T_{\rm eff} = 7594\pm18$ K, surface gravity log$g$ $=\ 3.93\pm0.09$ and metallicity [Fe/H] $=\ 0.12\pm0.02$.
However, the wide line profiles from HORuS high-resolution spectra show that J06163552 is a fast-rotating star, which means the parameter values from LAMOST LRS may be unreliable.

We determined $T_\mathrm{eff}$ and $\log g$ with the aid of grids of theoretical spectra,
which were created with iSpec \citep{Blanco2014,Blanco2019}, adopting the radial transfer code MOOG \citep{Sneden1973}
and the Kurucz/Castelli model atmosphere \citep{Kurucz1993,Castelli2003}.
By minimizing the value of $\chi^{2}$, we derived the most similar theoretical spectrum to the observed one and adopted its $T_\mathrm{eff}$ and $\log g$ as the parameters of J06163552.
We obtained the same parameters from the GTC/HORuS high-resolution and LAMOST med-resolution spectra, i.e. $T_\mathrm{eff} = 7900\pm200\,K$ and $\log g = 4.30\pm0.2$. The associated uncertainties were estimated considering the parameters' ranges covered by the 30 theoretical spectra with the smallest $\chi^{2}$ values.
Moreover, the Mg\,I/Mg\,II lines (seen in the HORuS spectra) reach ionization equilibrium considering Non-LTE effect, also supporting the derived $T_\mathrm{eff}$ and $\log g$ values.
The HORuS high-resolution spectra also constrain the $vsini$ of J06163552 to be $95\pm5$ km/s.

The {\it Gaia} DR2 gives a parallax of $\varpi = 0.9024 \pm 0.0271$ mas \citep{2018gaia}, corresponding to a distance of 1.11$\pm0.03$ kpc.
The $E(B-V)$ value is about 0.3, calculated with $E(B-V) = 0.884 \times {\rm (Bayestar19)}$, the latter\footnote{http://argonaut.skymaps.info/usage} of which ($\approx$0.342) is derived from the Pan-STARRS DR1 dust map \citep{2015ApJ...810...25G}.

\subsection{mass determination}
\label{pmass.sec}

\subsubsection{evolutionary mass estimation}
\label{subsubsect:mass_evo}

We used the ``{\it isochrones}" Python module \citep{2015ascl.soft03010M} to estimate stellar mass, which is an interpolation tool for the fitting of stellar models to photometric or spectroscopic parameters.
By using the trilinear interpolation in mass–age–[Fe/H] space for any given set of model grids, it is  able to predict physical or photometric properties provided by the models \citep{2015ApJ...809...25M}.
The input of the code includes the measured temperature, surface gravity, metallicity, multi-band magnitudes ($G$, $G_{\rm BP}$, $G_{\rm RP}$, $J$, $H$, and $K_{\rm S}$), {\it Gaia} DR2 parallax \citep{2018gaia} and extinction $A_V$ ($= 3.1 \times E(B-V)$).
The fitted mass and radius are:
$M\ =\ 1.69^{+0.14}_{-0.12}$ M$_{\odot}$ and $R\ =\ 1.65^{+0.27}_{-0.16}$ R$_{\odot}$
(Table \ref{parameters.tab}).

\begin{figure*}[!htbp]
   \center
   \includegraphics[width=0.98\textwidth]{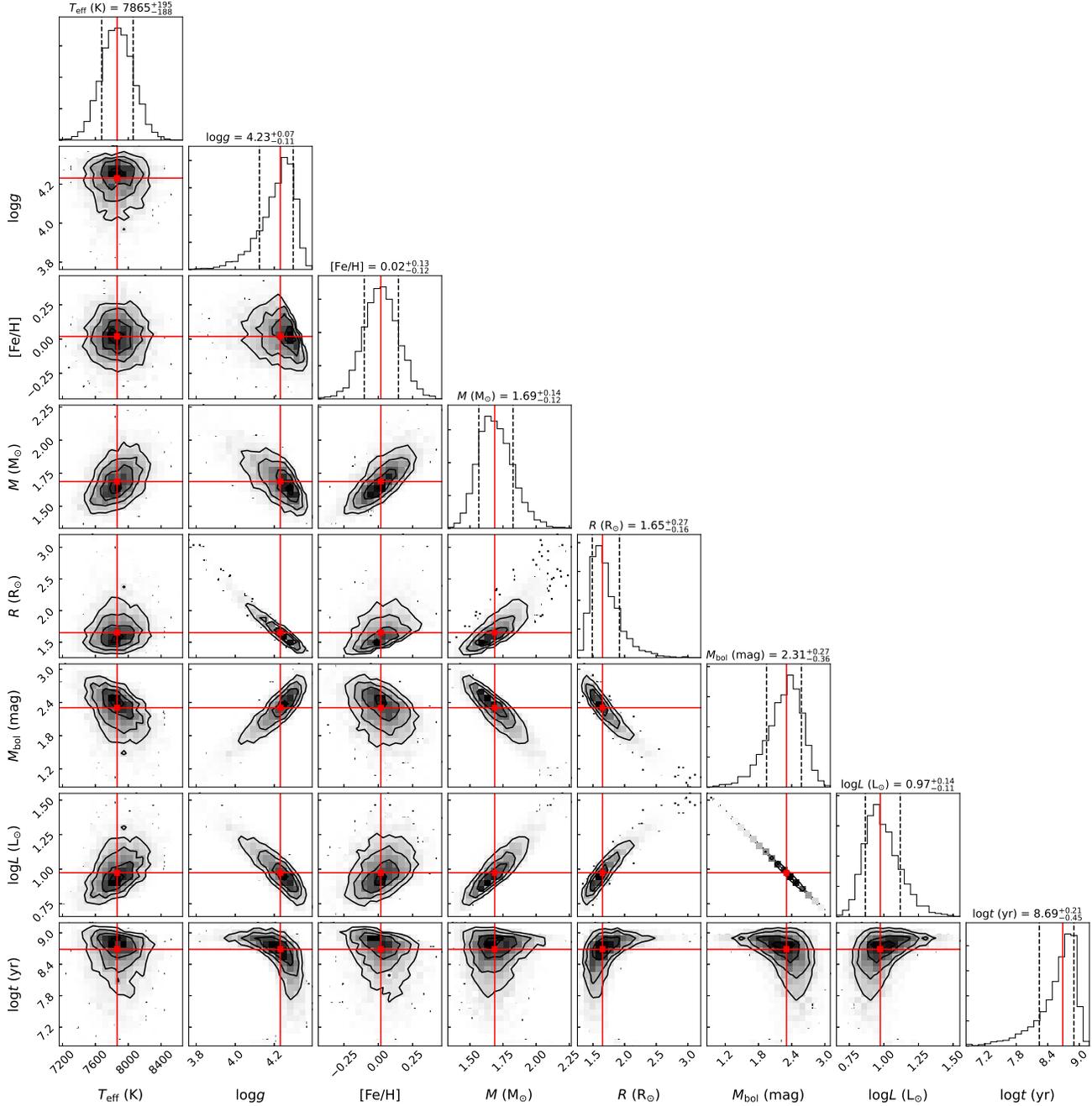}
   \caption{Corner plot showing the distribution of observed and physical parameters, derived from the {\it isochrones} code. The parameters are labeled as, effective temperature ($T_{\rm eff}$, in K), surface gravity (log$g$, in dex), metallicity ([Fe/H], in dex), mass ($M$, in M$_{\odot}$),  radius ($R$, in R$_{\odot}$), bolometric magnitude ($M_{\rm bol}$, in mag), bolometric luminosity (log$L$, in L$_{\odot}$), and age (log$t$, in yr).}
   \label{mistcorner.fig}
\end{figure*}

\subsubsection{Spectroscopic mass estimation}
\label{subsubsect:mass_sp}

The stellar mass can be estimated with the observed spectroscopic and photometric parameters.
First, we calculated the averaged bolometric magnitude by using the multi-band magnitudes ($G$, $G_{\rm BP}$, $G_{\rm RP}$, $J$, $H$, and $K_{\rm S}$), the {\it Gaia} DR2 distance, the extinction, and the bolometric corrections \citep[BC;][]{2019AA...632A.105C}.
For 2MASS magnitudes, we derived the attenuation by directly multiplying the extinction coefficients\footnote{http://argonaut.skymaps.info/usage} by the Bayestar19 value.
For {\it Gaia} magnitudes, we derived the extinction by multiplying $E(B-V)$ by the extinction coefficients from \citet{2018MNRAS.479L.102C}.
The BC for each band is derived from the PARSEC database\footnote{http://stev.oapd.inaf.it/YBC/}, with the input of $T_{\rm eff}$, log$g$, and [Fe/H] values.
Second, the bolometric luminosity of the visible star was calculated with the bolometric luminosity and and magnitude of the sun ($L_{\odot} =$ 3.83$\times$ 10$^{33}$ erg/s; $M_{\odot} =$ 4.74).
Finally, we derived the stellar mass with the bolometric luminosity, effective temperature, and surface gravity following
\begin{equation}
M = \frac{L}{4\pi~G\sigma~T^{4}}g.
\end{equation}
The spectroscopic mass is about 1.69$\pm$0.18 M$_{\odot}$, consistent with the evolutionary mass estimate.

\section{Orbital solution}
\label{orbit.sec}

\subsection{Radial Velocity fitting}

\begin{figure}[htbp!]
\center
\includegraphics[width=0.48\textwidth]{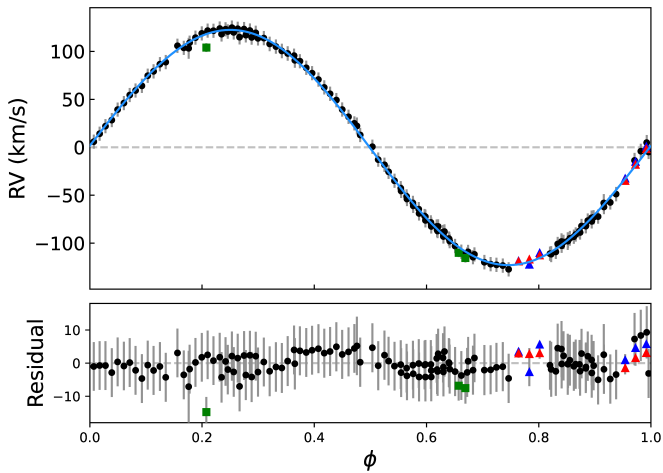}
\caption{Folded RV curves and the best binary orbital fit. The RV data are determined from the LAMOST low-resolution spectra (black circles), tue LAMOST med-resolution $B$- and $R$-band spectra (blue and red triangles, respectively), and the HORuS high-resolution spectra (green squares).}
\label{rv.fig}
\end{figure}

\begin{figure*}[htbp!]
\center
\includegraphics[width=1\textwidth]{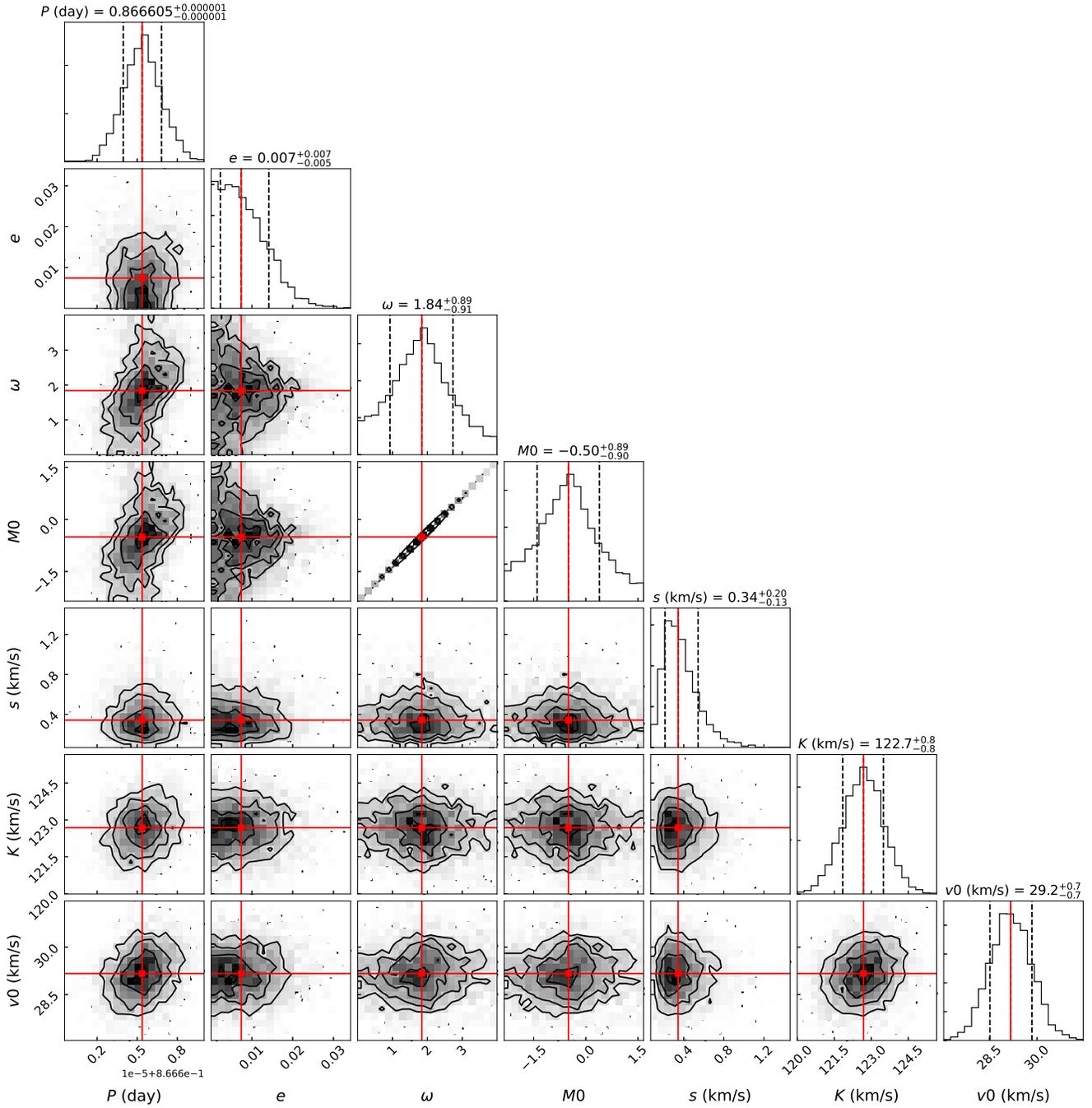}
\caption{Corner plot showing distribution of  orbital parameters derived from {\it The Joker}. The parameters are labeled as, orbital period ($P$, in days),  eccentricity of the system ($e$), argument of pericenter ($\omega$, in radians), mean anomaly at reference time ($M$0, in radians), extra ``jitter" added in quadrature to each visit-velocity error ($s$, in km/s), radial velocity semi-amplitude of the star ($K$,in km/s), and the center of mass velocity ($\nu$0,in km/s).}
\label{jokermcmc.fig}
\end{figure*}

We performed a Keplerian fit using the custom Markov chain Monte Carlo sampler {\it The Joker} \citep{2017ApJ...837...20P}.
{\it The Joker} works well with non-uniform data and allows to identify circular or eccentric orbits.
Both the LAMOST and HORuS data were used in the fitting.
Figure \ref{rv.fig} shows the RV data and orbital curve.
The fitted orbital parameters are:
period $P = 0.866605^{+0.000001}_{-0.000001}$ day,
eccentricity $e = 0.007^{+0.007}_{-0.005}$,
argument of the periastron $\omega = 1.84^{+0.89}_{-0.91}$,
mean anomaly at the first exposure $M_{\rm 0} = -0.50^{+0.89}_{-0.90}$,
semi-amplitude $K = 122.7^{+0.8}_{-0.8}$ km/s,
and systematic RV $\nu{\rm 0} = 29.2^{+0.7}_{-0.7}$ km/s (Table \ref{parameters.tab}).
Figure \ref{jokermcmc.fig} plots the MCMC results.

We derived the binary mass function $f$($M$) using posterior samples from our radial velocity modeling,
\begin{equation}
    f(M) = \frac{M_{2} \, \textrm{sin}^3 i} {(1+q)^{2}} = \frac{P \, K_{1}^{3} \, (1-e^2)^{3/2}}{2\pi G},
\end{equation}
\noindent
where $M_{1}$ is the mass of the visible star, $q = M_{1}/M_{2}$ is the mass ratio, and $i$ is the system inclination. Using the fit results, $f(M) \approx 0.169^{+0.001}_{-0.001}$ M$_{\odot}$, which represents a lower limit on the secondary mass ($M_2$ $\approx 1.06$ M$_{\odot}$ assuming $i = 90^{\circ}$).
Normal stars of this minimum mass would be apparent in the optical spectra; because it is unseen, we suspected that the secondary is a compact object.

\subsection{LC fitting with ELC}
\label{elc.sec}

The LCs of the visible star of J06163552 are dominated by ellipsoidal variations with an amplitude of $\approx$0.1 mag, and the deeper minima is fainter than the shallower minima by $\approx$0.01 mag, due to the limb-darkening and gravity darkening effects (Table \ref{lcvar.tab}).

The multi-band ($B$, $V$, $R$, and $I$) LCs from the observations by the 85-cm telescope can be used to constrain the mass ratio $q$ and inclination angle $i$, and finally determine the mass of the compact object.
We used the eclipsing LC code  \citep[ELC;][]{2000A&A...364..265O} to model the LC and RV.
Again, since there is no double-line feature in the optical spectra and no X-ray detection (Section \ref{star.uvxray}),
we assumed J06163552 contains a compact invisible secondary with no accretion disk and a tidally distorted companion.
The free parameters include the effective temperature of the visible companion $T_1$ (7700 K -- 8100 K), the mass of the visible star $M_1$ (1.5--1.8 M$_{\odot}$), the inclination angle $i$ (50$^{\circ}$--90$^{\circ}$), and the Roche lobe filling factor of the companion $f_1$ (0.1--1).
The period, mass function, eccentricity, K-velocity of the visible star were fixed as 0.8666 day, 0.167 M$_{\odot}$, 0, and 123 km/s, respectively.

\begin{figure*}[htbp!   ]
\center
\includegraphics[width=1\textwidth]{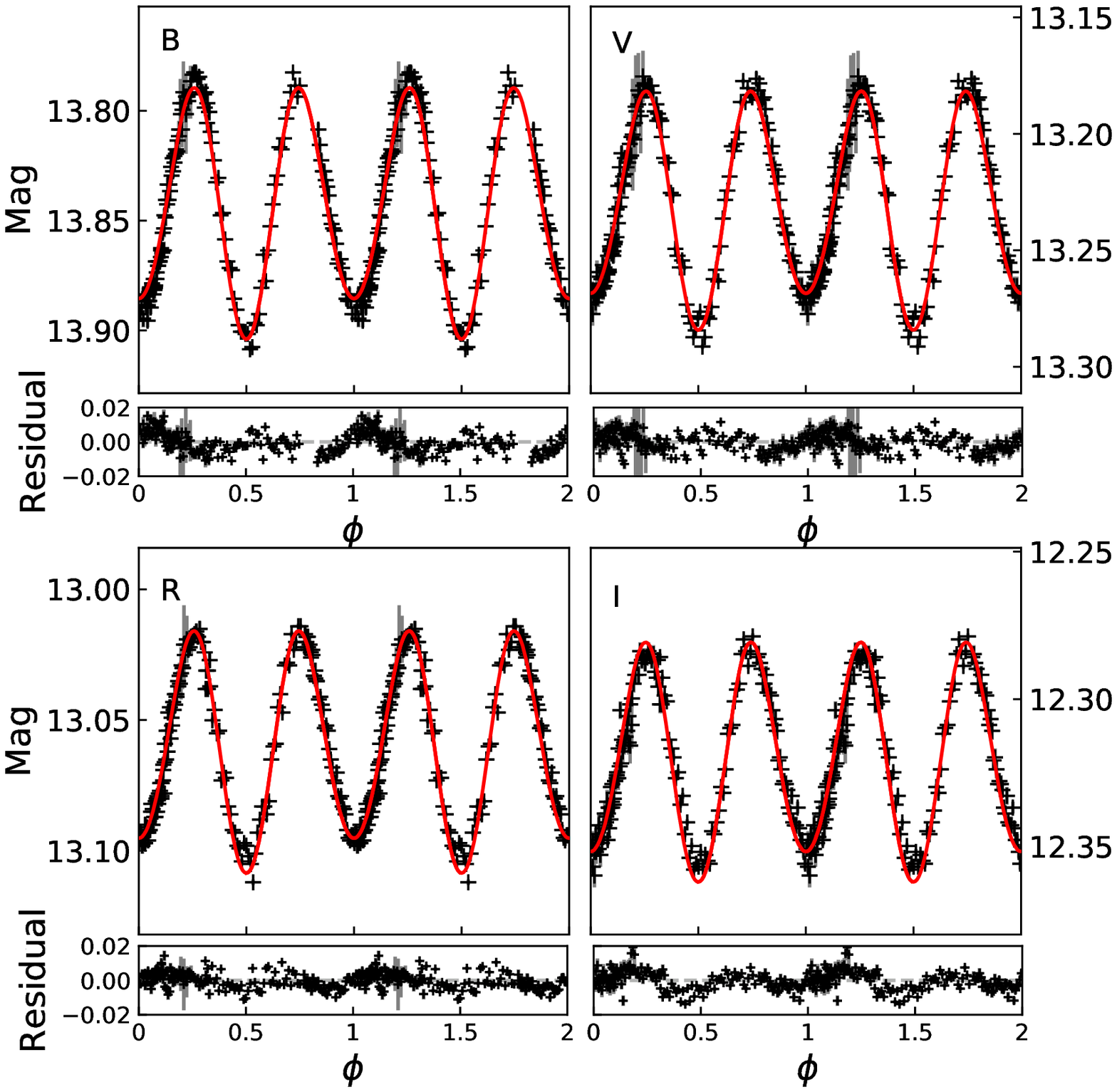}
\caption{Folded LCs in $B$, $V$, $R$, and $I$ filters. The red lines are from ELC modeling.}
\label{lc.fig}
\end{figure*}

The fitted results are $T_1 = 7735^{+149}_{-19}$ K, $M_1$ $=1.65^{+0.03}_{-0.04}$ M$_{\odot}$, $i =$ $77^{+3}_{-6}$ deg, and $f_1$ $=$ 0.67$^{+0.03}_{-0.01}$.
These corresponds to a mass ratio of $q = M_1/M_2 \approx$ 1.48 and a mass of the compact object of $M_2$ $\approx$ 1.11  M$_{\odot}$.
The fitted $f_1$ suggests the visible star hasn't filled it's Roche lobe.

\subsection{LC fitting with WD}
\label{wd.sec}

The Wilson-Devinney binary star modeling code \citep[WD;][]{1971ApJ...166..605W} is also used to model the LCs for comparison.
The visible star is identified as star 1.
The invisible compact star (star 2) is modeled by setting a very large potential ($\Omega_{\rm 2}$),
which will ensure an extremely small radius,
and an equal surface temperature to simplify the bolometric correction in the reflection effect,
as suggested by the WD tutorial.
The free parameters include
the mass ratio $q=(M_1/M_2)$
(note WD code actually uses $M_2/M_1$),
the inclination angle $INC$,
semi-major axis $SMA$,
and the potential of the visible star $\Omega_{\rm 1}$.
The luminosity $L_1$ was allowed to be scaled since we used relative flux for the four bands (i.e., $B$, $V$, $R$, and $I$).
The temperature of the visible star is fixed to be 7900 K following
high-resolution spectroscopic analysis (Section \ref{para.sec}).
The limb darken coefficients are internally determined by WD following the square root law.
Both the bolometric albedos for reflection $ALB$
and the gravity darkening exponent $GR$ are fixed to 1,
since the adopted temperature indicates a radiative envelope.

\begin{figure*}[htbp!]
\center
\includegraphics[width=1\textwidth]{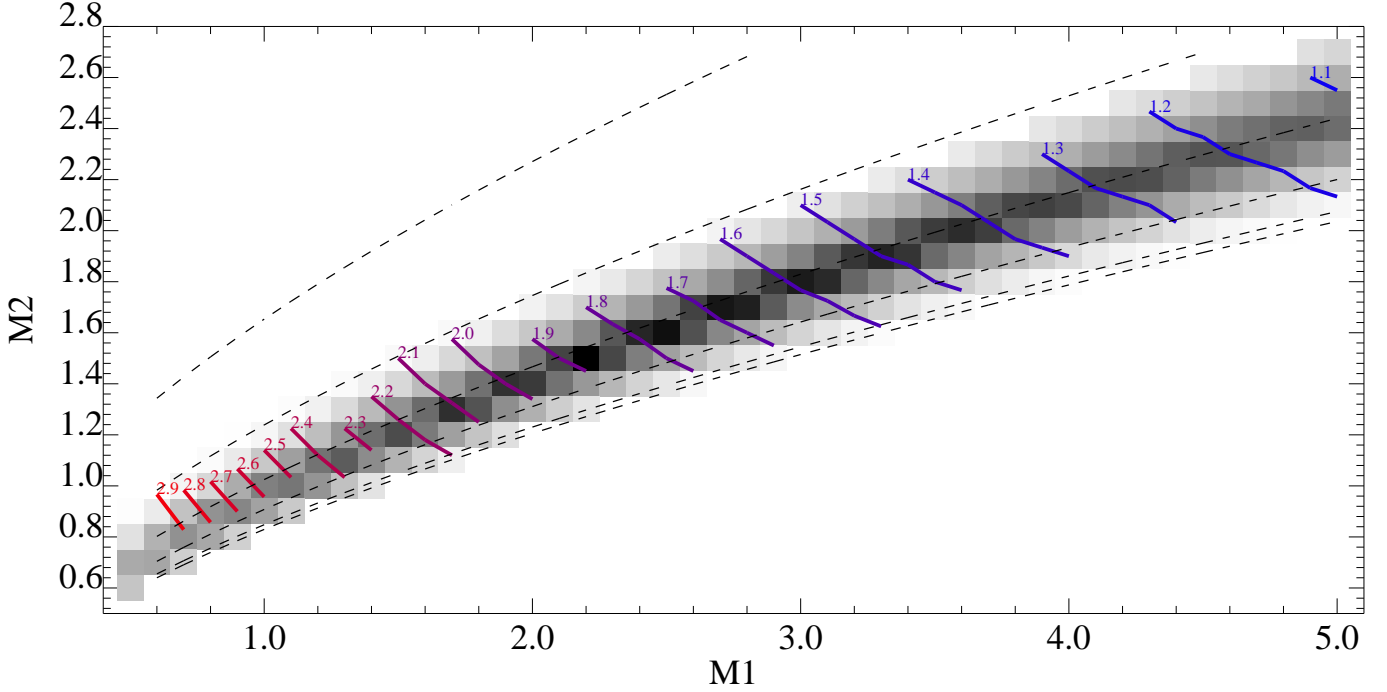}
\caption{Light curve fit on Grid of $M_1$ and $M_2$ with the WD program.
The boxes with darker color denote smaller $\chi^2$.
The dashed lines denote equal inclination angles,
which are 90$^{\circ}$, 80$^{\circ}$, 70$^{\circ}$, 
60$^{\circ}$, 50$^{\circ}$ and 40$^{\circ}$, from bottom to top.
The solid lines connect equal $M_{\rm bol}$ points.
The $M_{\rm bol}$ estimated with GAIA data is $\approx$2.15 mag.}
\label{wdm1m2}
\end{figure*}

To better constrain the model with RV observations,
a 2D grid search is applied instead of fitting with WD directly.
The first dimension is the mass of visible component ($M_1$),
from 0.5 to 5.0 M$_{\odot}$, with a step of 0.1 M$_{\odot}$.
The second dimension is the mass of the compact invisible one ($M_2$).
The range of $M_2$ is set to be 0.3 to 1.0 M$_{\odot}$ initially but adjusted as the iteration goes on,
in order to properly cover the best $M_2$ value; the step is fixed to be 0.1 M$_{\odot}$.
For each $M_1$ and $M_2$ point, the $SMA$, $q$, and $INC$ is directly calculated and fixed in WD,
based on orbital period and semi-amplitude of RV of the visible star,
leaving the radius $R_1$ or $\Omega_{\rm 1}$ and $L_1$ to be adjusted by the WD programs.
The result is shown in Figure \ref{wdm1m2}.

As we can see for each $M_1$, one best $M_2$ can be estimated.
However, the difference between the best fits, corresponding to different $q$ and $INC$ values,
cannot be distinguished by eyes.
The reduced $\chi^2$ of these fits are also very close, 
which means the $q$ can not be well constrained by current LC and RV data.
Nevertheless, the calculated bolometric magnitude $M_{\rm bol}$ from model can be used as an additional constraint,
as shown in Figure \ref{wdm1m2}.
With the GAIA $G$-band magnitude (13.21 mag),
distance (1.11 kpc),
bolometric correction (0.06 mag) and $G$-band extinction ($\approx$0.90 mag),
as discussed in Section \ref{pmass.sec},
the bolometric magnitude can be estimated to be 2.15$\pm$0.10 mag.
The value indicates that the best fit point is at
$M_1$ of $\approx1.6\pm0.1$ M$_{\odot}$
and $M_2$ of $\approx1.3\pm0.1$ M$_{\odot}$.
The $M_1$ value agrees well with the mass estimation discussed in Section \ref{pmass.sec}.
The model also produces log$g$ $=$ 4.15, $SMA=$ 5.45 R$_{\odot}$,
and radius $=1.77$ R$_{\odot}$ for the visible component.

\begin{figure}[htbp!]
\center
\includegraphics[width=0.50\textwidth]{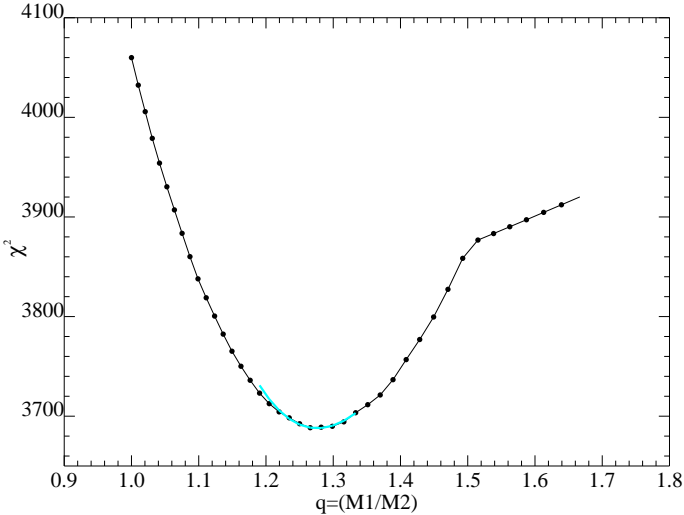}
\caption{
The mass ratio $q=M_1/M_2$ grid search with $M_1=1.6$ M$_{\odot}$.
The estimated $q$ is $1.27\pm0.01$,
corresponding to $M_2=1.26\pm0.01$ M$_{\odot}$.
}
\label{wdqgrid}
\end{figure}

Finally, by fixing $M_1=1.6$ M$_{\odot}$,
a finer $q$ grid search was carried out, as shown in Figure \ref{wdqgrid}.
The best fit $M_2$ is $\approx$ 1.26$\pm$0.01 M$_{\odot}$,
corresponding to
$q (M_1/M_2)=1.27\pm0.01$,
$INC=62.2\pm1^{\circ}$,
$SMA=5.43\pm0.01$ R$_{\odot}$,
and $\Omega_{\rm 1}=3.93\pm0.01$.
Other best-fit parameters include
$R_1=1.76$ R$_{\odot}$,
$log {\rm g} =4.15$
and $M_{\rm bol}=2.16$ mag.
Besides, if $M_1$ is chosen to be $1.69$ M$_{\odot}$, according to Section \ref{pmass.sec},
the best fit $M_2$ is $\approx$ 1.30$\pm$0.01 M$_{\odot}$,
corresponding to  $M_{\rm bol}\approx2.11$.
The filling factor of the visible star,
calculated by dividing the radius with the Roche lobe volume radius \citep{1989ApJ...347..998E}, is about 70\%.

Assuming the orbital and rotational motions are synchronized, together with the orbital period, stellar radius ($\approx1.7$ R$_{\odot}$) and inclination angle ($60^{\circ}$--$80^{\circ}$), the $vsini$ of the visible star can be estimated to be 89--101 km/s, which is consistent with the estimation from HORuS spectra. This suggests that our estimation of the inclination angle, and thus the mass of the secondary, are reasonable.

\section{Discussion}
\label{diss.sec}

\begin{table*}
\caption{Parameter Summaries for J06163552.
\label{parameters.tab}}
\setlength{\tabcolsep}{4.5pt}
\begin{center}
 \begin{tabular}{cccc}
\hline\noalign{\smallskip}
Parameter & Unit & Value & From \\
\hline\noalign{\smallskip}
RA & [hh:mm:ss.ss] & 06:16:35.52 & \multirow{2}*{UCAC4 Catalogue} \\
DEC & [dd:mm:ss.s] & 23:19:09.3 &  \\
$V$ mag & [mag] & 13.23$\pm$0.04 & \citet{2013AJ....145...44Z} \\
Distance & [kpc] & 1.11$\pm0.03$ & \citet{2018gaia} \\
$E(B-V)$ & [mag] & 0.3$\pm$0.01 & \citet{2015ApJ...810...25G} \\
\hline\noalign{\smallskip}
$T_{\rm eff,1}$ & [K] & $7900\pm200$ & \multirow{3}*{HORuS spectra} \\
log $g_{\rm 1}$ & [dex] & $4.3\pm0.2$ &  \\
$vsini_{\rm 1}$ & [km/s] & $95\pm5$ & \\
\noalign{\smallskip}\hline
 $T_{\rm eff,1}$ & [K] & $7865^{+195}_{-188}$ & \multirow{8}*{{\it isochrones} fitting}  \\
 log $g_{\rm 1}$ & [dex] & $4.23^{+0.07}_{-0.11}$ &  \\
 $\left[{\rm Fe/H}\right]_{\rm 1}$ & [dex] & $0.02^{+0.13}_{-0.12}$ &  \\
 $M_{\rm 1}$ & [$M_{\odot}$] & $1.69^{+0.14}_{-0.12}$ &  \\
 $R_{\rm 1}$ & [$R_{\odot}$] & $1.65^{+0.27}_{-0.16}$ &  \\
 $M_{\rm bol,1}$ & [mag] & $2.31^{+0.27}_{-0.36}$ &  \\
 log $L_{\rm bol,1}$ & [$L_{\odot}$] & $0.97^{+0.14}_{-0.11}$ &  \\
\noalign{\smallskip}\hline
$P$ & [day] & $0.866605^{+0.000001}_{-0.000001}$ & \multirow{6}*{{\it The Joker} fitting}  \\
$e$ & [-] & $0.007^{+0.007}_{-0.005}$ &  \\
$\omega$ & [radians] & $1.84^{+0.89}_{-0.91}$ & \\
$M_{\rm 0}$ & [radians] & $-0.50^{+0.89}_{-0.90}$ &  \\
$K$ & [km/s] & $122.7^{+0.8}_{-0.8}$ &  \\
$\nu_{\rm 0}$ & [km/s] & $29.2^{+0.7}_{-0.7}$ &  \\
\noalign{\smallskip}\hline
 $T_{\rm eff,1}$ & [K] & $7735^{+149}_{-19}$ & \multirow{5}*{ELC fitting} \\
 $M_{\rm 1}$ & [$M_{\odot}$] & $1.65^{+0.03}_{-0.04}$ &  \\
inclination angle $i$ & [deg] & $77^{+3}_{-6}$ &  \\
filling factor $f_{\rm 1}$ & [-] & 0.67$^{+0.03}_{-0.01}$ & \\
 $M_{\rm 2}$ & [$M_{\odot}$] & $\approx1.11$ &  \\
\noalign{\smallskip}\hline
inclination angle $INC$ & [deg] & 62$\pm$5 & \multirow{6}*{WD fitting} \\
Semi-major axis $SMA$ & [$R_{\odot}$] & $5.45\pm0.10$ &  \\
 $M_{\rm 1}$ & [$M_{\odot}$] & $1.60\pm0.10$ &  \\
 $R_{\rm 1}$ & [$R_{\odot}$] & $1.76\pm0.10$ &  \\
log $g_{\rm 1}$ & [dex] & $4.15\pm0.10$ &  \\
$M_{\rm 2}$ & [$M_{\odot}$] & $1.30\pm0.10$ &  \\
\noalign{\smallskip}\hline
\end{tabular}
\end{center}
\smallskip
\end{table*}

\begin{table*}
\caption{Light curve features for J06163552.
\label{lcvar.tab}}
\setlength{\tabcolsep}{4.5pt}
\begin{center}
 \begin{tabular}{cccc}
\hline\noalign{\smallskip}
Band & Median magnitude & Amplitude & Minima Difference \\
 & [mag] & [mag] & [mag] \\
\hline\noalign{\smallskip}
$B$  & 13.843          & 0.116$\pm$0.005 & 0.015$\pm$0.005  \\
$V$ & 13.23            & 0.102$\pm$0.007 & 0.013$\pm$0.007  \\
$R$  & 13.06           & 0.083$\pm$0.007 & 0.006$\pm$0.006  \\
$I$  & 12.32           & 0.069$\pm$0.004 & 0.003$\pm$0.006  \\
ZTF $g$   & 13.56      & 0.111$\pm$0.021 & 0.021$\pm$0.018 \\
ZTF $r$   & 13.21      & 0.099$\pm$0.029 & 0.022$\pm$0.039 \\
ASAS-SN $V$  & 13.27   & 0.090$\pm$0.015 & 0.013$\pm$0.022 \\
\noalign{\smallskip}\hline
\end{tabular}
\end{center}
{The maxima, deeper minima and shallower minima of the LCs were calculated as averaged values with about 10 data points around the phases (0.25, 0.75), (0, 1.0) and (0.5), respectively.}
\end{table*}

\subsection{The Nature of the undetected object}

The undetected component could be a compact object or a main-sequence star.
In the latter case (i.e., normal binary), by using the $M$-$L$ relation $M$ $\propto$ $L^{4.3}$ for stars with masses between 1.05 M$_{\odot}$ and 2.4 M$_{\odot}$ \citep{2018MNRAS.479.5491E}, the luminosity ratio of the two stars is 3--7. 
Double-line feature should be seen in the HORuS spectra.
Especially, the HORuS observation times are close to the phase of 0.25 and 0.75,
at which the two stars moving backwards or towards to us.
The line separation of the two normal stars can be calculated as $\Delta \lambda = K/c\times\lambda$, which means at 5000 \AA\ the lines of the visible star shifts about 2 \AA\ and the lines of the other star shifts about 2.7 \AA\ (using the averaged mass ratio estimated from the ELC and WD code) towards opposite direction.
As shown in Figure \ref{spfig}, we compared the three HORuS high-resolution spectra and one spectral template with the stellar parameters of the visible star ($T_{\rm eff}=7900$ K, log$g=4.2$ and [Fe/H] $=$ 0). No same absorption line appears in these residuals, suggesting J06163552 is a single-line spectroscopic binary.

\begin{figure*}[htbp!]
\center
\includegraphics[width=1\textwidth]{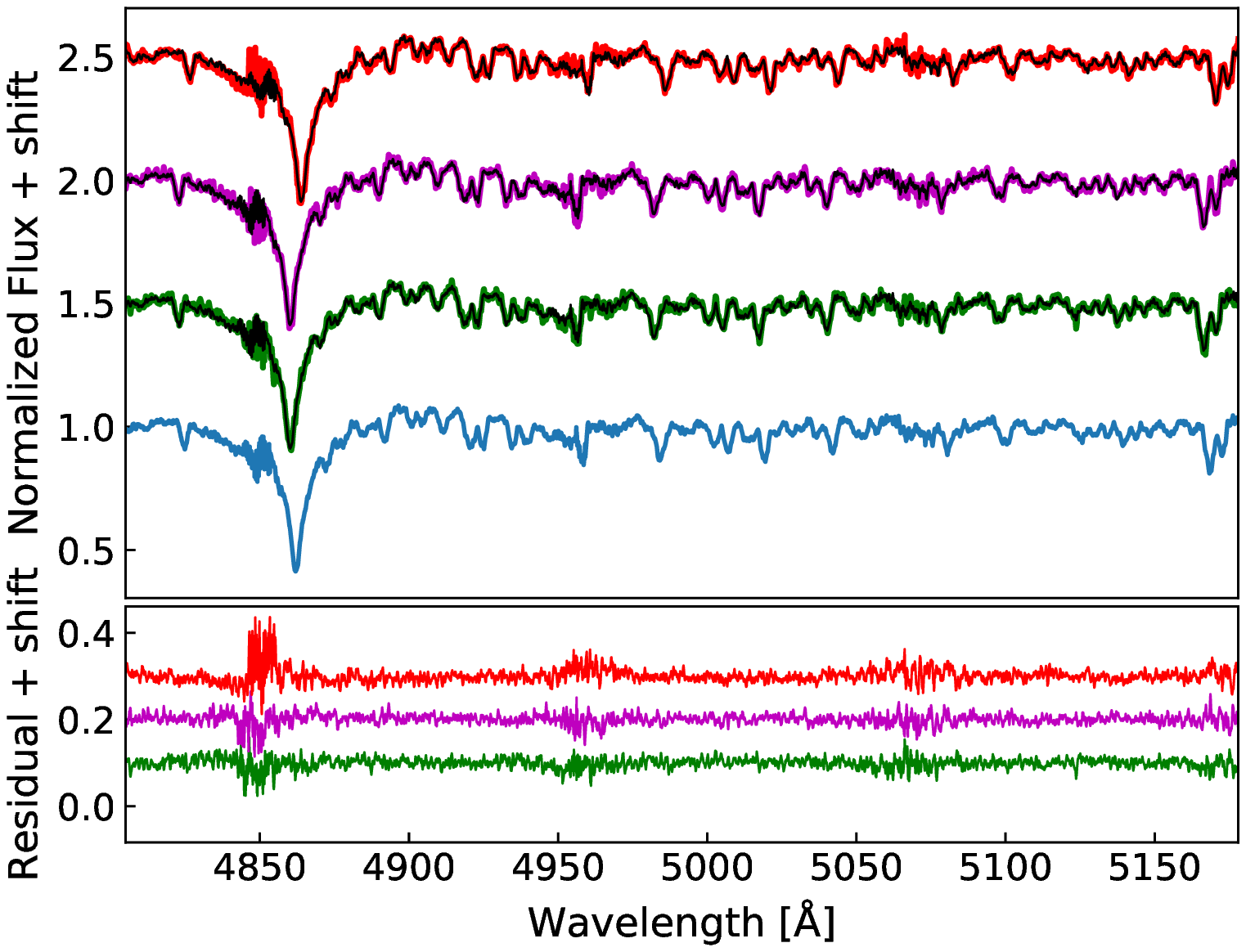}
\caption{
Top panel: Model spectra produced by FDBINARY assuming J06163552 is a single-line binary. The lower blue line is the disentangled spectrum in the rest frame. The upper three lines (from bottom to top) are observational HORuS spectra (BMJD $=$ 58809.25208, 58809.26249, and 58814.06205, respectively), overlapped with the disentangled spectrum (black lines) doppler-shifted with the fitted RV by FDBINARY. 
Bottom panel: Residuals showing the difference between the observed spectra and disentangled spectrum. No clear signal of absorption lines can be seen from the residuals. These bumps with large residual are the boundaries of different spectral orders.}
\label{specall}
\end{figure*}

We further used the Fourier domain-based disentangling code FDBinary \citep[][new version: fd3\footnote{http://sail.zpf.fer.hr/fdbinary/}]{2004ASPC..318..111I} to perform spectral disentangling.
The used spectral region ranges from 4800 ${\rm \AA}$ to 5200 ${\rm \AA}$.
We tried two modes by treating J06163552 is a single-line binary or double-line binary, respectively. 
If J06163552 is a normal binary, the residual of the single-line binary mode would show clear absorption features, while the double-line binary mode would produce two stellar spectra.
The period and eccentricity are fixed to be 0.8666 days and 0.
The semi-amplitude of RV of star 1 ($K_{\rm 1}$) is set to be 123.3 km/s, varying in the range of $\pm$5 km/s.
In the double-line binary case, the semi-amplitude of star 2 ($K_{\rm 2}$) varies in the range of 135--215 km/s.
In both the two modes, no clear signal of the other component was detected, further indicating the companion is a compact object (i.e., WD or NS).
Figure \ref{specall} shows the spectral disentangling assuming J06163552 is a single-line binary.

We did some tests to check the detection limit of the secondary star with this method for our source. Assuming a mass range of 0.7--1.3 M$_{\odot}$, we derived approximate effective temperature and surface gravity of the secondary (assuming a main-sequence star) for each mass value (0.7, 0.8, 0.9, 1.0, 1.1, 1.2, and 1.3 M$_{\odot}$)\footnote{http://www.pas.rochester.edu/\~emamajek/EEM\_dwarf\_UBV-
IJHK\_colors\_Teff.txt}.
We picked out model templates from Phoenix\footnote{https://phoenix.astro.physik.uni-goettingen.de}, reduced their resolution to $R=$ 24500, and applied rotational broadening ($v{\rm sin}i =$ 50 km/s, 100 km/s and 150 km/s). 
Then the model spectra and the observational spectra were combined following \citep{2022arXiv220706996K}
\begin{equation}
    {f}_{\lambda,{\rm binary}}=\frac{{f}_{\lambda,2} + k_\lambda {f}_{\lambda,1}}{1+k_\lambda},
	\label{eq:bolzmann}
\end{equation}
 where  $k_\lambda= \frac{B_\lambda(T{_{\rm eff,1}})~M_1}{B_\lambda(T{_{\rm eff,2}})~M_2} 10^{{\rm log}g_2-{\rm log}g_1}$ is the luminosity ratio per wavelength unit, $B_\lambda$ is the black-body radiation  (Plank function), and $M$ is the mass.
We ran the FDBinary code with the grid of faked binary spectra (4800 ${\rm \AA}$ to 5200 ${\rm \AA}$), and the spectral disentangling results by visual check are listed in Table \ref{disentangling.tab}. 
When the secondary mass is higher than 1 M$_{\odot}$, this method can well separate the binary components; when the secondary is lower than 0.8 M$_{\odot}$, the spectra can't be disentangled, mostly due to the low SNR of the faked spectra.
Considering that the secondary mass of our source is between 1.1 to 1.3 M$_{\odot}$, the disentangling method is suitable for our target.

In addition, we tried to use the ELC code to find a normal binary solution. The free parameters include the inclination angle ($i$), the Roche lobe filling factor of the visible star ($f_{\rm 1}$), the temperatures of the two stars ($T_{\rm 1}$ and $T_{\rm 2}$), and the mass ratio ($q$).
The fitting results are $T_{\rm 1}$ $\approx$ 7700 K and $T_{\rm 2}$ $\approx$ 4700 K. However, the inclination angle ($i$ $\approx$ 56$^o$) and mass ratio ($q$ $\approx$ 0.9) led to mass estimates of $M_{\rm 1}$ $\approx$ 1.4 M$_{\odot}$ and $M_{\rm 2}$ $\approx$ 1.3 M$_{\odot}$. 
Considering that the visible star hasn't filling its roche lobe (Section \ref{elc.sec} and \ref{rochelobe.sec}), the temperature and mass of the cooler star are incompatible, again suggesting  J06163552 isn't a normal binary.

\begin{table}
\caption{Spectral disentangling results by visual check. "\ding{52}\ding{52}" means the spectra are well disentangled, "\ding{52}" means the spectra can be disentangled but in low significance, and "\ding{53}" means the spectra can not be disentangled. \label{disentangling.tab}}
\centering
\setlength{\tabcolsep}{4pt}
 \begin{tabular}{cccc}
\hline\noalign{\smallskip}
 $M_{\rm 2}$ & Type & $v{\rm sin}i$ & 
\begin{tabular}{c}Disentangling  \\ results \end{tabular} \\
\hline\noalign{\smallskip}
\multirow{3}*{1.3} & \multirow{3}*{F5V} &50 & \ding{52}\ding{52}\\
   &   & 100 & \ding{52}\ding{52}\\
   &    & 150 & \ding{52}\\
   \cline{1-4}
\multirow{3}*{1.2}& \multirow{3}*{F7V} & 50 & \ding{52}\ding{52}\\
    &  & 100 & \ding{52}\ding{52}\\
    &   & 150 & \ding{52}\\
   \cline{1-4}
   \multirow{3}*{1.1}&\multirow{3}*{F9V}  & 50 & \ding{52}\ding{52}\\
    &  & 100 & \ding{52}\ding{52}\\
    &   & 150 & \ding{52}\\
   \cline{1-4}
   \multirow{3}*{1.0} &\multirow{3}*{G2V} & 50 & \ding{52}\ding{52}\\
    &  & 100 & \ding{52}\ding{52}\\
    &   & 150 & \ding{53}\\
   \cline{1-4}
   \multirow{3}*{0.9}& \multirow{3}*{G5V} & 50 & \ding{52}\ding{52}\\
    &  & 100 & \ding{52}\\
    &   & 150 & \ding{53}\\
   \cline{1-4}
   \multirow{3}*{0.8} &\multirow{3}*{K2V} & 50 & \ding{52}\ding{52}\\
   &   & 100 & \ding{53}\\
    &   & 150 & \ding{53}\\
   \cline{1-4}
      \multirow{3}*{0.7}& \multirow{3}*{K5V} & 50 & \ding{53}\\
  &   &  100 & \ding{53}\\
   &    & 150 & \ding{53}\\
\noalign{\smallskip}\hline
\end{tabular}
\end{table}

\subsection{No detection of UV and X-ray emission}
\label{star.uvxray}

ROSAT consists one Wide Field Camera covering the wavelength ranges of 60--140 \AA\ and 110--200 \AA, and one X-ray telescope, on which the instrument Position Sensitive Proportional Counter performed a nearly 100\% sky survey.
The sky field of J06163552 has been covered by ROSAT during 1990--1991, but without any detection of the object
\citep[][]{1995AdSpR..16c..69M,2016A&A...588A.103B}.
The sky field of J06163552 has been covered by the PSPC pointed observation of IC 443 (i.e., Jellyfish Nebula) for five times, i.e., ObsID rp500020n00 (exposure of $\sim$3168 sec), rp500021n00 ($\sim$4011 sec), rp500022n00 ($\sim$2912 sec), rp500045a00 ($\sim$1099 sec), and rp500045a01 ($\sim$7002 sec).
No X-ray source was found at the location of J06163552.
We used the ESA upper limit server\footnote{http://xmmuls.esac.esa.int/upperlimitserver/} to derive upper limits of X-ray emission.
Although it shows that one object was possibly detected with a count rate of 0.0273$\pm$0.0116 in one observation (ObsID rp500021n00), we suspected that it is due to the various and strong background from visual check.
Using these upper limit estimations, we obtained an averaged 3$\sigma$ upper limit of $\sim$0.018 count/sec, corresponding to a flux of $2.15\times 10^{-13} \rm erg\, s^{-1}\, cm^{-2}$ in the 0.2--2 keV range (assuming $N_H = $ 1$\times10^{21}$ cm$^{-2}$ and $\tau =$ 2 for a power-law spectrum).
At a distance $d=1.11\,$kpc (Section \ref{para.sec}), this corresponds to a luminosity of $L_X$ $\lesssim$ 3.17$\times$10$^{31}$ $\rm erg\, s^{-1}$.
In addition, the sky field of J06163552 was also covered by the PSPC sky survey, with an exposure time of 355 sec. This gives an upper limit of 4.67 $\times 10^{-13} \rm erg\, s^{-1}\, cm^{-2}$ for flux and 6.90$\times$10$^{31}$ $\rm erg\, s^{-1}$ for luminosity in the 0.2–2 keV range.
We also find no UV signal basing on the ROSAT WFC images.
Therefore, It is possible J06163552 has no or very weak UV and X-ray emission.
This may support that J06163552 is a low mass NS candidate rather than a (young hot) white dwarf.
Future follow up with UV observation \citep[e.g., CSST/NUV;][]{2018MNRAS.480.2178C} or radio observation may help pin down the nature of the companion.

\subsection{Roche lobe of the visible star}
\label{rochelobe.sec}

Both the ELC and WD fittings return a small Roche-lobe filling factor of $\approx$0.7 (Section \ref{elc.sec} and \ref{wd.sec}), suggesting that no accretion disk is present or no accretion exists. This is consistent with that we didn't detect any X-ray or UV signal or emission feature in any spectrum.
In addition, if one star fills it's Roche lobe, an accurate measurement of the rotational broadening could provide a valuable independent check on the mass ratio of the binary \citep{1986ApJ...304..371G}.
In that case, the companion star has been tidal locked with the binary orbit, the rotational broadening is
$vsini = (K_1 + K_2)R_2/a = 0.462 K_1 q^{1/3}(1+q)^{2/3}$, where $q = M_1/M_2$ \citep{1988ApJ...324..411W}.
The $vsini$ value of J06163552 ($\approx$95 km/s) indicates a mass ratio of $\approx$ 1.08 and a mass of 1.56 M$_{\odot}$ for the compact star.
This is much higher than our estimations, which also suggests J06163552 has not filled its Roche lobe.

\section{Summary}
\label{sum.sec}

We present the discovery of J06163552 by using LAMOST LRS data, which is a binary with orbital period $P =$ 0.8666 day containing an early-type star and a possible neutron star.
The system’s optical LCs (from observations by ASAS-SN, ZTF, and 85-cm telescope in Xinglong) are dominated by ellipsoidal variability.
From the HORuS high-resolution spectra, we derive the effective temperature $T_{\rm eff}=7900\pm200$ K and surface gravity log$g$ $=4.3\pm0.1$, suggesting a late A-type star with mass of $\approx$ 1.65 M$_{\odot}$.
By fitting the multi-band LCs and RV curve, we constrain the physical properties of the system, including the period, RV semi-amplitude, inclination angle, and mass ratio, etc.
The mass of the unseen object is estimated as $\approx$ 1.1--1.3 M$_{\odot}$.
No double-line feature is seen from the HORuS spectra, and spectral disentangling shows no additional component with absorption spectra, supporting that the binary system J06163552 contains a compact component.
No X-ray or UV signal is found in the ROSAT archive data, suggests J06163552 includes one neutron star, rather than a (young hot) white dwarf.

LAMOST has started a new 5-year survey program, LAMOST II, containing both TD and non-TD surveys \citep[see][, for more details]{2020arXiv200507210L, 2020ApJS..251...15Z}. By using the low- and med-resolution TD survey of four $K$2 plates, \citet{2021RAA....21..292W} present a binary catalog ($\sim$ 2700 objects), and some binaries with  compact components are under study.
The LAMOST TD survey and similar spectroscopic campaigns will help increase the sample of compact objects, a quiescent population different from the X-ray-bright one, and test the BH/NS formation rate in binaries.
Many binary population synthesis calculations focus on binaries containing BHs and normal-star companions.
\citet{2019ApJ...885..151S} reported that there are thousands of detached BH binaries in the Milky Way, with hundreds of them being potentially observable systems. Most detached BH binaries have main-sequence companions and only a small percent of them have giant companions.
By convolving the visibility of BH binaries with the LAMOST detection sensitivity, \citet{2019ApJ...886...97Y} predicted that about 400 candidate BH binaries can be found by the non-TD low-resolution survey, while about 50--350 candidates can be detected by the med-resolution TD survey.
Several possible NSs or WDs have been identified with the LAMOST spectroscopic data \citep[e.g.,][]{2021MNRAS.505.2051E, 2022arXiv220611270M, 2022arXiv220900765X, 2022NatAs.tmp..201Y}.
We expect that more binaries including compact objects will be discovered by the LAMOST survey.

\section*{acknowledgements}
We thank the anonymous referee for helpful comments and suggestions that have  improved the paper. 
Guoshoujing Telescope (the Large Sky Area Multi-Object Fiber Spectroscopic Telescope LAMOST) is a National Major Scientific Project built by the Chinese Academy of Sciences. Funding for the project has been provided by the National Development and Reform Commission. LAMOST is operated and managed by the National Astronomical Observatories, Chinese Academy of Sciences.
This work presents results from the European Space Agency (ESA) space mission Gaia. Gaia data are being processed by the Gaia Data Processing and Analysis Consortium (DPAC). Funding for the DPAC is provided by national institutions, in particular the institutions participating in the Gaia MultiLateral Agreement (MLA). The Gaia mission website is https://www.cosmos.esa.int/gaia. The Gaia archive website is https://archives.esac.esa.int/gaia.
We acknowledge use of the VizieR catalogue access tool, operated at CDS, Strasbourg, France, and of Astropy, a community-developed core Python package for Astronomy \citep[Astropy; ][]{2022ApJ...935..167A}.
This research made use of Photutils \citep{larry_bradley_2020_4044744}, an Astropy package for detection and photometry of astronomical sources.
This work was supported by National Science Foundation of China (NSFC) under grant numbers 11933004/11988101/12090041/12273056/12273057, National Key Research and Development Program of China (NKRDPC) under grant numbers 2019YFA0405000 and 2019YFA0405504, and Strategic Priority Program of the Chinese Academy of Sciences under grant number XDB41000000.
H.-L. Y. and S. W. acknowledges support from the Youth Innovation Promotion Association of the CAS (id. 2020060 and 2019057, respectively).
H.-L. Y. and Z.-R. B. acknowledges support from the Cultivation Project for LAMOST Scientific Payoff and Research Achievement of CAMS-CAS.

\software{
MOOG \citep{Sneden1973},
ATLAS9 \citep[][]{Castelli2003},
ELC \citep{2000A&A...364..265O},
Wilson-Devinney binary star modeling code \citep[][]{1971ApJ...166..605W}, 
FDBinary \citep[][]{2004ASPC..318..111I}, 
The Joker \citep{2017ApJ...837...20P}, 
isochrones \citep{2015ascl.soft03010M}.}

\bibliographystyle{aasjournal}
\bibliography{bibtex.bib}{}

\begin{thebibliography}{}
\expandafter\ifx\csname natexlab\endcsname\relax\def\natexlab#1{#1}\fi
\providecommand{\url}[1]{\href{#1}{#1}}
\providecommand{\dodoi}[1]{doi:~\href{http://doi.org/#1}{\nolinkurl{#1}}}
\providecommand{\doeprint}[1]{\href{http://ascl.net/#1}{\nolinkurl{http://ascl.net/#1}}}
\providecommand{\doarXiv}[1]{\href{https://arxiv.org/abs/#1}{\nolinkurl{https://arxiv.org/abs/#1}}}

\bibitem[{{Abbott} {et~al.}(2017){Abbott}, {Abbott}, {Abbott}, {Acernese},
  {Ackley}, {Adams}, {Adams}, {Addesso}, {Adhikari}, {Adya}, {Affeldt},
  {Afrough}, {Agarwal}, {Agathos}, {Agatsuma}, {Aggarwal}, {Aguiar}, {Aiello},
  {Ain}, {Ajith}, {Allen}, {Allen}, {Allocca}, {Altin}, {Amato}, {Ananyeva},
  {Anderson}, {Anderson}, {Angelova}, {Antier}, {Appert}, {Arai}, {Araya},
  {Areeda}, {Arnaud}, {Arun}, {Ascenzi}, {Ashton}, {Ast}, {Aston}, {Astone},
  {Atallah}, {Aufmuth}, {Aulbert}, {AultONeal}, {Austin}, {Avila-Alvarez},
  {Babak}, {Bacon}, {Bader}, {Bae}, {Bailes}, {Baker}, {Baldaccini},
  {Ballardin}, {Ballmer}, {Banagiri}, {Barayoga}, {Barclay}, {Barish},
  {Barker}, {Barkett}, {Barone}, {Barr}, {Barsotti}, {Barsuglia}, {Barta},
  {Barthelmy}, {Bartlett}, {Bartos}, {Bassiri}, {Basti}, {Batch}, {Bawaj},
  {Bayley}, {Bazzan}, {B{\'e}csy}, {Beer}, {Bejger}, {Belahcene}, {Bell},
  {Berger}, {Bergmann}, {Bernuzzi}, {Bero}, {Berry}, {Bersanetti}, {Bertolini},
  {Betzwieser}, {Bhagwat}, {Bhandare}, {Bilenko}, {Billingsley}, {Billman},
  {Birch}, {Birney}, {Birnholtz}, {Biscans}, {Biscoveanu}, {Bisht}, {Bitossi},
  {Biwer}, {Bizouard}, {Blackburn}, {Blackman}, {Blair}, {Blair}, {Blair},
  {Bloemen}, {Bock}, {Bode}, {Boer}, {Bogaert}, {Bohe}, {Bondu}, {Bonilla},
  {Bonnand}, {Boom}, {Bork}, {Boschi}, {Bose}, {Bossie}, {Bouffanais}, {Bozzi},
  {Bradaschia}, {Brady}, {Branchesi}, {Brau}, {Briant}, {Brillet}, {Brinkmann},
  {Brisson}, {Brockill}, {Broida}, {Brooks}, {Brown}, {Brown}, {Brunett},
  {Buchanan}, {Buikema}, {Bulik}, {Bulten}, {Buonanno}, {Buskulic}, {Buy},
  {Byer}, {Cabero}, {Cadonati}, {Cagnoli}, {Cahillane}, {Calder{\'o}n
  Bustillo}, {Callister}, {Calloni}, {Camp}, {Canepa}, {Canizares}, {Cannon},
  {Cao}, {Cao}, {Capano}, {Capocasa}, {Carbognani}, {Caride}, {Carney},
  {Carullo}, {Casanueva Diaz}, {Casentini}, {Caudill}, {Cavagli{\`a}},
  {Cavalier}, {Cavalieri}, {Cella}, {Cepeda}, {Cerd{\'a}-Dur{\'a}n},
  {Cerretani}, {Cesarini}, {Chamberlin}, {Chan}, {Chao}, {Charlton}, {Chase},
  {Chassande-Mottin}, {Chatterjee}, {Chatziioannou}, {Cheeseboro}, {Chen},
  {Chen}, {Chen}, {Cheng}, {Chia}, {Chincarini}, {Chiummo}, {Chmiel}, {Cho},
  {Cho}, {Chow}, {Christensen}, {Chu}, {Chua}, {Chua}, {Chung}, {Chung},
  {Ciani}, {Ciolfi}, {Cirelli}, {Cirone}, {Clara}, {Clark}, {Clearwater},
  {Cleva}, {Cocchieri}, {Coccia}, {Cohadon}, {Cohen}, {Colla}, {Collette},
  {Cominsky}, {Constancio}, {Conti}, {Cooper}, {Corban}, {Corbitt},
  {Cordero-Carri{\'o}n}, {Corley}, {Cornish}, {Corsi}, {Cortese}, {Costa},
  {Coughlin}, {Coughlin}, {Coulon}, {Countryman}, {Couvares}, {Covas}, {Cowan},
  {Coward}, {Cowart}, {Coyne}, {Coyne}, {Creighton}, {Creighton}, {Cripe},
  {Crowder}, {Cullen}, {Cumming}, {Cunningham}, {Cuoco}, {Dal Canton},
  {D{\'a}lya}, {Danilishin}, {D'Antonio}, {Danzmann}, {Dasgupta}, {Da Silva
  Costa}, {Dattilo}, {Dave}, {Davier}, {Davis}, {Daw}, {Day}, {De}, {DeBra},
  {Degallaix}, {De Laurentis}, {Del{\'e}glise}, {Del Pozzo}, {Demos}, {Denker},
  {Dent}, {De Pietri}, {Dergachev}, {De Rosa}, {DeRosa}, {De Rossi}, {DeSalvo},
  {de Varona}, {Devenson}, {Dhurandhar}, {D{\'\i}az}, {Dietrich}, {Di Fiore},
  {Di Giovanni}, {Di Girolamo}, {Di Lieto}, {Di Pace}, {Di Palma}, {Di Renzo},
  {Doctor}, {Dolique}, {Donovan}, {Dooley}, {Doravari}, {Dorrington},
  {Douglas}, {Dovale {\'A}lvarez}, {Downes}, {Drago}, {Dreissigacker},
  {Driggers}, {Du}, {Ducrot}, {Dudi}, {Dupej}, {Dwyer}, {Edo}, {Edwards},
  {Effler}, {Eggenstein}, {Ehrens}, {Eichholz}, {Eikenberry}, {Eisenstein},
  {Essick}, {Estevez}, {Etienne}, {Etzel}, {Evans}, {Evans}, {Factourovich},
  {Fafone}, {Fair}, {Fairhurst}, {Fan}, {Farinon}, {Farr}, {Farr},
  {Fauchon-Jones}, {Favata}, {Fays}, {Fee}, {Fehrmann}, {Feicht}, {Fejer},
  {Fernandez-Galiana}, {Ferrante}, {Ferreira}, {Ferrini}, {Fidecaro},
  {Finstad}, {Fiori}, {Fiorucci}, {Fishbach}, {Fisher}, {Fitz-Axen},
  {Flaminio}, {Fletcher}, {Fong}, {Font}, {Forsyth}, {Forsyth}, {Fournier},
  {Frasca}, {Frasconi}, {Frei}, {Freise}, {Frey}, {Frey}, {Fries}, {Fritschel},
  {Frolov}, {Fulda}, {Fyffe}, {Gabbard}, {Gadre}, {Gaebel}, {Gair},
  {Gammaitoni}, {Ganija}, {Gaonkar}, {Garcia-Quiros}, {Garufi}, {Gateley},
  {Gaudio}, {Gaur}, {Gayathri}, {Gehrels}, {Gemme}, {Genin}, {Gennai},
  {George}, {George}, {Gergely}, {Germain}, {Ghonge}, {Ghosh}, {Ghosh},
  {Ghosh}, {Giaime}, {Giardina}, {Giazotto}, {Gill}, {Glover}, {Goetz},
  {Goetz}, {Gomes}, {Goncharov}, {Gonz{\'a}lez}, {Gonzalez Castro},
  {Gopakumar}, {Gorodetsky}, {Gossan}, {Gosselin}, {Gouaty}, {Grado}, {Graef},
  {Granata}, {Grant}, {Gras}, {Gray}, {Greco}, {Green}, {Gretarsson}, {Groot},
  {Grote}, {Grunewald}, {Gruning}, {Guidi}, {Guo}, {Gupta}, {Gupta}, {Gushwa},
  {Gustafson}, {Gustafson}, {Halim}, {Hall}, {Hall}, {Hamilton}, {Hammond},
  {Haney}, {Hanke}, {Hanks}, {Hanna}, {Hannam}, {Hannuksela}, {Hanson},
  {Hardwick}, {Harms}, {Harry}, {Harry}, {Hart}, {Haster}, {Haughian}, {Healy},
  {Heidmann}, {Heintze}, {Heitmann}, {Hello}, {Hemming}, {Hendry}, {Heng},
  {Hennig}, {Heptonstall}, {Heurs}, {Hild}, {Hinderer}, {Ho}, {Hoak}, {Hofman},
  {Holt}, {Holz}, {Hopkins}, {Horst}, {Hough}, {Houston}, {Howell}, {Hreibi},
  {Hu}, {Huerta}, {Huet}, {Hughey}, {Husa}, {Huttner}, {Huynh-Dinh}, {Indik},
  {Inta}, {Intini}, {Isa}, {Isac}, {Isi}, {Iyer}, {Izumi}, {Jacqmin}, {Jani},
  {Jaranowski}, {Jawahar}, {Jim{\'e}nez-Forteza}, {Johnson},
  {Johnson-McDaniel}, {Jones}, {Jones}, {Jonker}, {Ju}, {Junker}, {Kalaghatgi},
  {Kalogera}, {Kamai}, {Kandhasamy}, {Kang}, {Kanner}, {Kapadia}, {Karki},
  {Karvinen}, {Kasprzack}, {Kastaun}, {Katolik}, {Katsavounidis}, {Katzman},
  {Kaufer}, {Kawabe}, {K{\'e}f{\'e}lian}, {Keitel}, {Kemball}, {Kennedy},
  {Kent}, {Key}, {Khalili}, {Khan}, {Khan}, {Khan}, {Khazanov}, {Kijbunchoo},
  {Kim}, {Kim}, {Kim}, {Kim}, {Kim}, {Kim}, {Kimbrell}, {King}, {King},
  {Kinley-Hanlon}, {Kirchhoff}, {Kissel}, {Kleybolte}, {Klimenko}, {Knowles},
  {Koch}, {Koehlenbeck}, {Koley}, {Kondrashov}, {Kontos}, {Korobko}, {Korth},
  {Kowalska}, {Kozak}, {Kr{\"a}mer}, {Kringel}, {Krishnan}, {Kr{\'o}lak},
  {Kuehn}, {Kumar}, {Kumar}, {Kumar}, {Kuo}, {Kutynia}, {Kwang}, {Lackey},
  {Lai}, {Landry}, {Lang}, {Lange}, {Lantz}, {Lanza}, {Larson},
  {Lartaux-Vollard}, {Lasky}, {Laxen}, {Lazzarini}, {Lazzaro}, {Leaci},
  {Leavey}, {Lee}, {Lee}, {Lee}, {Lee}, {Lee}, {Lehmann}, {Lenon}, {Leon},
  {Leonardi}, {Leroy}, {Letendre}, {Levin}, {Li}, {Linker}, {Littenberg},
  {Liu}, {Liu}, {Lo}, {Lockerbie}, {London}, {Lord}, {Lorenzini}, {Loriette},
  {Lormand}, {Losurdo}, {Lough}, {Lousto}, {Lovelace}, {L{\"u}ck}, {Lumaca},
  {Lundgren}, {Lynch}, {Ma}, {Macas}, {Macfoy}, {Machenschalk}, {MacInnis},
  {Macleod}, {Maga{\~n}a Hernandez}, {Maga{\~n}a-Sandoval}, {Maga{\~n}a
  Zertuche}, {Magee}, {Majorana}, {Maksimovic}, {Man}, {Mandic}, {Mangano},
  {Mansell}, {Manske}, {Mantovani}, {Marchesoni}, {Marion}, {M{\'a}rka},
  {M{\'a}rka}, {Markakis}, {Markosyan}, {Markowitz}, {Maros}, {Marquina},
  {Marsh}, {Martelli}, {Martellini}, {Martin}, {Martin}, {Martynov}, {Marx},
  {Mason}, {Massera}, {Masserot}, {Massinger}, {Masso-Reid}, {Mastrogiovanni},
  {Matas}, {Matichard}, {Matone}, {Mavalvala}, {Mazumder}, {McCarthy},
  {McClelland}, {McCormick}, {McCuller}, {McGuire}, {McIntyre}, {McIver},
  {McManus}, {McNeill}, {McRae}, {McWilliams}, {Meacher}, {Meadors}, {Mehmet},
  {Meidam}, {Mejuto-Villa}, {Melatos}, {Mendell}, {Mercer}, {Merilh},
  {Merzougui}, {Meshkov}, {Messenger}, {Messick}, {Metzdorff}, {Meyers},
  {Miao}, {Michel}, {Middleton}, {Mikhailov}, {Milano}, {Miller}, {Miller},
  {Miller}, {Millhouse}, {Milovich-Goff}, {Minazzoli}, {Minenkov}, {Ming},
  {Mishra}, {Mitra}, {Mitrofanov}, {Mitselmakher}, {Mittleman}, {Moffa},
  {Moggi}, {Mogushi}, {Mohan}, {Mohapatra}, {Molina}, {Montani}, {Moore},
  {Moraru}, {Moreno}, {Morisaki}, {Morriss}, {Mours}, {Mow-Lowry}, {Mueller},
  {Muir}, {Mukherjee}, {Mukherjee}, {Mukherjee}, {Mukund}, {Mullavey}, {Munch},
  {Mu{\~n}iz}, {Muratore}, {Murray}, {Nagar}, {Napier}, {Nardecchia},
  {Naticchioni}, {Nayak}, {Neilson}, {Nelemans}, {Nelson}, {Nery}, {Neunzert},
  {Nevin}, {Newport}, {Newton}, {Ng}, {Nguyen}, {Nguyen}, {Nichols}, {Nielsen},
  {Nissanke}, {Nitz}, {Noack}, {Nocera}, {Nolting}, {North}, {Nuttall},
  {Oberling}, {O'Dea}, {Ogin}, {Oh}, {Oh}, {Ohme}, {Okada}, {Oliver},
  {Oppermann}, {Oram}, {O'Reilly}, {Ormiston}, {Ortega}, {O'Shaughnessy},
  {Ossokine}, {Ottaway}, {Overmier}, {Owen}, {Pace}, {Page}, {Page}, {Pai},
  {Pai}, {Palamos}, {Palashov}, {Palomba}, {Pal-Singh}, {Pan}, {Pan}, {Pang},
  {Pang}, {Pankow}, {Pannarale}, {Pant}, {Paoletti}, {Paoli}, {Papa}, {Parida},
  {Parker}, {Pascucci}, {Pasqualetti}, {Passaquieti}, {Passuello}, {Patil},
  {Patricelli}, {Pearlstone}, {Pedraza}, {Pedurand}, {Pekowsky}, {Pele},
  {Penn}, {Perez}, {Perreca}, {Perri}, {Pfeiffer}, {Phelps}, {Piccinni},
  {Pichot}, {Piergiovanni}, {Pierro}, {Pillant}, {Pinard}, {Pinto}, {Pirello},
  {Pitkin}, {Poe}, {Poggiani}, {Popolizio}, {Porter}, {Post}, {Powell},
  {Prasad}, {Pratt}, {Pratten}, {Predoi}, {Prestegard}, {Prijatelj},
  {Principe}, {Privitera}, {Prix}, {Prodi}, {Prokhorov}, {Puncken}, {Punturo},
  {Puppo}, {P{\"u}rrer}, {Qi}, {Quetschke}, {Quintero}, {Quitzow-James},
  {Raab}, {Rabeling}, {Radkins}, {Raffai}, {Raja}, {Rajan}, {Rajbhandari},
  {Rakhmanov}, {Ramirez}, {Ramos-Buades}, {Rapagnani}, {Raymond}, {Razzano},
  {Read}, {Regimbau}, {Rei}, {Reid}, {Reitze}, {Ren}, {Reyes}, {Ricci},
  {Ricker}, {Rieger}, {Riles}, {Rizzo}, {Robertson}, {Robie}, {Robinet},
  {Rocchi}, {Rolland}, {Rollins}, {Roma}, {Romano}, {Romano}, {Romel}, {Romie},
  {Rosi{\'n}ska}, {Ross}, {Rowan}, {R{\"u}diger}, {Ruggi}, {Rutins}, {Ryan},
  {Sachdev}, {Sadecki}, {Sadeghian}, {Sakellariadou}, {Salconi}, {Saleem},
  {Salemi}, {Samajdar}, {Sammut}, {Sampson}, {Sanchez}, {Sanchez},
  {Sanchis-Gual}, {Sandberg}, {Sanders}, {Sassolas}, {Sathyaprakash},
  {Saulson}, {Sauter}, {Savage}, {Sawadsky}, {Schale}, {Scheel}, {Scheuer},
  {Schmidt}, {Schmidt}, {Schnabel}, {Schofield}, {Sch{\"o}nbeck}, {Schreiber},
  {Schuette}, {Schulte}, {Schutz}, {Schwalbe}, {Scott}, {Scott}, {Seidel},
  {Sellers}, {Sengupta}, {Sentenac}, {Sequino}, {Sergeev}, {Shaddock},
  {Shaffer}, {Shah}, {Shahriar}, {Shaner}, {Shao}, {Shapiro}, {Shawhan},
  {Sheperd}, {Shoemaker}, {Shoemaker}, {Siellez}, {Siemens}, {Sieniawska},
  {Sigg}, {Silva}, {Singer}, {Singh}, {Singhal}, {Sintes}, {Slagmolen},
  {Smith}, {Smith}, {Smith}, {Somala}, {Son}, {Sonnenberg}, {Sorazu},
  {Sorrentino}, {Souradeep}, {Spencer}, {Srivastava}, {Staats}, {Staley},
  {Steinke}, {Steinlechner}, {Steinlechner}, {Steinmeyer}, {Stevenson},
  {Stone}, {Stops}, {Strain}, {Stratta}, {Strigin}, {Strunk}, {Sturani},
  {Stuver}, {Summerscales}, {Sun}, {Sunil}, {Suresh}, {Sutton}, {Swinkels},
  {Szczepa{\'n}czyk}, {Tacca}, {Tait}, {Talbot}, {Talukder}, {Tanner},
  {T{\'a}pai}, {Taracchini}, {Tasson}, {Taylor}, {Taylor}, {Tewari}, {Theeg},
  {Thies}, {Thomas}, {Thomas}, {Thomas}, {Thorne}, {Thorne}, {Thrane},
  {Tiwari}, {Tiwari}, {Tokmakov}, {Toland}, {Tonelli}, {Tornasi},
  {Torres-Forn{\'e}}, {Torrie}, {T{\"o}yr{\"a}}, {Travasso}, {Traylor},
  {Trinastic}, {Tringali}, {Trozzo}, {Tsang}, {Tse}, {Tso}, {Tsukada}, {Tsuna},
  {Tuyenbayev}, {Ueno}, {Ugolini}, {Unnikrishnan}, {Urban}, {Usman},
  {Vahlbruch}, {Vajente}, {Valdes}, {Vallisneri}, {van Bakel}, {van Beuzekom},
  {van den Brand}, {Van Den Broeck}, {Vander-Hyde}, {van der Schaaf}, {van
  Heijningen}, {van Veggel}, {Vardaro}, {Varma}, {Vass}, {Vas{\'u}th},
  {Vecchio}, {Vedovato}, {Veitch}, {Veitch}, {Venkateswara}, {Venugopalan},
  {Verkindt}, {Vetrano}, {Vicer{\'e}}, {Viets}, {Vinciguerra}, {Vine}, {Vinet},
  {Vitale}, {Vo}, {Vocca}, {Vorvick}, {Vyatchanin}, {Wade}, {Wade}, {Wade},
  {Walet}, {Walker}, {Wallace}, {Walsh}, {Wang}, {Wang}, {Wang}, {Wang},
  {Wang}, {Ward}, {Warner}, {Was}, {Watchi}, {Weaver}, {Wei}, {Weinert},
  {Weinstein}, {Weiss}, {Wen}, {Wessel}, {We{\ss}els}, {Westerweck},
  {Westphal}, {Wette}, {Whelan}, {Whitcomb}, {Whiting}, {Whittle}, {Wilken},
  {Williams}, {Williams}, {Williamson}, {Willis}, {Willke}, {Wimmer},
  {Winkler}, {Wipf}, {Wittel}, {Woan}, {Woehler}, {Wofford}, {Wong}, {Worden},
  {Wright}, {Wu}, {Wysocki}, {Xiao}, {Yamamoto}, {Yancey}, {Yang}, {Yap},
  {Yazback}, {Yu}, {Yu}, {Yvert}, {Zadro{\.Z}ny}, {Zanolin}, {Zelenova},
  {Zendri}, {Zevin}, {Zhang}, {Zhang}, {Zhang}, {Zhang}, {Zhao}, {Zhou},
  {Zhou}, {Zhu}, {Zhu}, {Zimmerman}, {Zucker}, {Zweizig}, {LIGO Scientific
  Collaboration}, \& {Virgo Collaboration}}]{2017PhRvL.119p1101A}
{Abbott}, B.~P., {Abbott}, R., {Abbott}, T.~D., {et~al.} 2017, \prl, 119,
  161101, \dodoi{10.1103/PhysRevLett.119.161101}

\bibitem[{{Antoniadis} {et~al.}(2013){Antoniadis}, {Freire}, {Wex}, {Tauris},
  {Lynch}, {van Kerkwijk}, {Kramer}, {Bassa}, {Dhillon}, {Driebe}, {Hessels},
  {Kaspi}, {Kondratiev}, {Langer}, {Marsh}, {McLaughlin}, {Pennucci}, {Ransom},
  {Stairs}, {van Leeuwen}, {Verbiest}, \& {Whelan}}]{2013Sci...340..448A}
{Antoniadis}, J., {Freire}, P. C.~C., {Wex}, N., {et~al.} 2013, Science, 340,
  448, \dodoi{10.1126/science.1233232}

\bibitem[{{Astropy Collaboration} {et~al.}(2022){Astropy Collaboration},
  {Price-Whelan}, {Lim}, {Earl}, {Starkman}, {Bradley}, {Shupe}, {Patil},
  {Corrales}, {Brasseur}, {N{\"o}the}, {Donath}, {Tollerud}, {Morris},
  {Ginsburg}, {Vaher}, {Weaver}, {Tocknell}, {Jamieson}, {van Kerkwijk},
  {Robitaille}, {Merry}, {Bachetti}, {G{\"u}nther}, {Aldcroft},
  {Alvarado-Montes}, {Archibald}, {B{\'o}di}, {Bapat}, {Barentsen},
  {Baz{\'a}n}, {Biswas}, {Boquien}, {Burke}, {Cara}, {Cara}, {Conroy},
  {Conseil}, {Craig}, {Cross}, {Cruz}, {D'Eugenio}, {Dencheva}, {Devillepoix},
  {Dietrich}, {Eigenbrot}, {Erben}, {Ferreira}, {Foreman-Mackey}, {Fox},
  {Freij}, {Garg}, {Geda}, {Glattly}, {Gondhalekar}, {Gordon}, {Grant},
  {Greenfield}, {Groener}, {Guest}, {Gurovich}, {Handberg}, {Hart},
  {Hatfield-Dodds}, {Homeier}, {Hosseinzadeh}, {Jenness}, {Jones}, {Joseph},
  {Kalmbach}, {Karamehmetoglu}, {Ka{\l}uszy{\'n}ski}, {Kelley}, {Kern},
  {Kerzendorf}, {Koch}, {Kulumani}, {Lee}, {Ly}, {Ma}, {MacBride}, {Maljaars},
  {Muna}, {Murphy}, {Norman}, {O'Steen}, {Oman}, {Pacifici}, {Pascual},
  {Pascual-Granado}, {Patil}, {Perren}, {Pickering}, {Rastogi}, {Roulston},
  {Ryan}, {Rykoff}, {Sabater}, {Sakurikar}, {Salgado}, {Sanghi}, {Saunders},
  {Savchenko}, {Schwardt}, {Seifert-Eckert}, {Shih}, {Jain}, {Shukla}, {Sick},
  {Simpson}, {Singanamalla}, {Singer}, {Singhal}, {Sinha}, {Sip{\H{o}}cz},
  {Spitler}, {Stansby}, {Streicher}, {{\v{S}}umak}, {Swinbank}, {Taranu},
  {Tewary}, {Tremblay}, {Val-Borro}, {Van Kooten}, {Vasovi{\'c}}, {Verma}, {de
  Miranda Cardoso}, {Williams}, {Wilson}, {Winkel}, {Wood-Vasey}, {Xue},
  {Yoachim}, {Zhang}, {Zonca}, \& {Astropy Project
  Contributors}}]{2022ApJ...935..167A}
{Astropy Collaboration}, {Price-Whelan}, A.~M., {Lim}, P.~L., {et~al.} 2022,
  \apj, 935, 167, \dodoi{10.3847/1538-4357/ac7c74}

\bibitem[{{Bai} {et~al.}(2017){Bai}, {Zhang}, {Yuan}, {Li}, {Chen}, {Lei},
  {Yang}, {Dong}, {Wang}, \& {Zhao}}]{2017PASP....17.91}
{Bai}, Z.-R., {Zhang}, H.-T., {Yuan}, H.-L., {et~al.} 2017, Research in
  Astronomy and Astrophysics, 17, 91

\bibitem[{{Bai} {et~al.}(2021){Bai}, {Zhang}, {Yuan}, {Fan}, {He}, {Lei},
  {Dong}, {Yu}, {Zhao}, {Zhang}, {Hou}, \& {Chu}}]{2021RAA....21..249B}
---. 2021, Research in Astronomy and Astrophysics, 21, 249,
  \dodoi{10.1088/1674-4527/21/10/249}

\bibitem[{{Blanco-Cuaresma}(2019)}]{Blanco2019}
{Blanco-Cuaresma}, S. 2019, \mnras, 486, 2075, \dodoi{10.1093/mnras/stz549}

\bibitem[{{Blanco-Cuaresma} {et~al.}(2014){Blanco-Cuaresma}, {Soubiran},
  {Heiter}, \& {Jofr{\'e}}}]{Blanco2014}
{Blanco-Cuaresma}, S., {Soubiran}, C., {Heiter}, U., \& {Jofr{\'e}}, P. 2014,
  \aap, 569, A111, \dodoi{10.1051/0004-6361/201423945}

\bibitem[{{Boller} {et~al.}(2016){Boller}, {Freyberg}, {Tr{\"u}mper}, {Haberl},
  {Voges}, \& {Nandra}}]{2016A&A...588A.103B}
{Boller}, T., {Freyberg}, M.~J., {Tr{\"u}mper}, J., {et~al.} 2016, \aap, 588,
  A103, \dodoi{10.1051/0004-6361/201525648}

\bibitem[{Bradley {et~al.}(2020)Bradley, Sipőcz, Robitaille, Tollerud,
  Vinícius, Deil, Barbary, Wilson, Busko, Günther, Cara, Conseil, Bostroem,
  Droettboom, Bray, Bratholm, Lim, Barentsen, Craig, Pascual, Perren, Greco,
  Donath, de~Val-Borro, Kerzendorf, Bach, Weaver, D'Eugenio, Souchereau, \&
  Ferreira}]{larry_bradley_2020_4044744}
Bradley, L., Sipőcz, B., Robitaille, T., {et~al.} 2020, astropy/photutils:
  1.0.0, 1.0.0,  Zenodo, \dodoi{10.5281/zenodo.4044744}

\bibitem[{{Cao} {et~al.}(2018){Cao}, {Gong}, {Meng}, {Xu}, {Chen}, {Guo}, {Li},
  {Liu}, {Xue}, {Cao}, {Fu}, {Zhang}, {Wang}, \& {Zhan}}]{2018MNRAS.480.2178C}
{Cao}, Y., {Gong}, Y., {Meng}, X.-M., {et~al.} 2018, \mnras, 480, 2178,
  \dodoi{10.1093/mnras/sty1980}

\bibitem[{{Casagrande} \& {VandenBerg}(2018)}]{2018MNRAS.479L.102C}
{Casagrande}, L., \& {VandenBerg}, D.~A. 2018, \mnras, 479, L102,
  \dodoi{10.1093/mnrasl/sly104}

\bibitem[{{Casares} {et~al.}(2014){Casares}, {Negueruela}, {Rib{\'o}}, {Ribas},
  {Paredes}, {Herrero}, \& {Sim{\'o}n-D{\'\i}az}}]{2014Natur.505..378C}
{Casares}, J., {Negueruela}, I., {Rib{\'o}}, M., {et~al.} 2014, \nat, 505, 378,
  \dodoi{10.1038/nature12916}

\bibitem[{{Castelli} \& {Kurucz}(2003)}]{Castelli2003}
{Castelli}, F., \& {Kurucz}, R.~L. 2003, in Modelling of Stellar Atmospheres,
  ed. N.~{Piskunov}, W.~W. {Weiss}, \& D.~F. {Gray}, Vol. 210, A20.
\newblock \doarXiv{astro-ph/0405087}

\bibitem[{{Chen} {et~al.}(2019){Chen}, {Girardi}, {Fu}, {Bressan}, {Aringer},
  {Dal Tio}, {Pastorelli}, {Marigo}, {Costa}, \& {Zhang}}]{2019AA...632A.105C}
{Chen}, Y., {Girardi}, L., {Fu}, X., {et~al.} 2019, \aap, 632, A105,
  \dodoi{10.1051/0004-6361/201936612}

\bibitem[{{Cromartie} {et~al.}(2020){Cromartie}, {Fonseca}, {Ransom},
  {Demorest}, {Arzoumanian}, {Blumer}, {Brook}, {DeCesar}, {Dolch}, {Ellis},
  {Ferdman}, {Ferrara}, {Garver-Daniels}, {Gentile}, {Jones}, {Lam}, {Lorimer},
  {Lynch}, {McLaughlin}, {Ng}, {Nice}, {Pennucci}, {Spiewak}, {Stairs},
  {Stovall}, {Swiggum}, \& {Zhu}}]{2020NatAs...4...72C}
{Cromartie}, H.~T., {Fonseca}, E., {Ransom}, S.~M., {et~al.} 2020, Nature
  Astronomy, 4, 72, \dodoi{10.1038/s41550-019-0880-2}

\bibitem[{{Cui} {et~al.}(2012){Cui}, {Zhao}, {Chu}, {Li}, {Li}, {Zhang}, {Su},
  {Yao}, {Wang}, {Xing}, {Li}, {Zhu}, {Wang}, {Gu}, {Luo}, {Xu}, {Zhang},
  {Liu}, {Zhang}, {Yang}, {Cao}, {Chen}, {Chen}, {Chen}, {Chen}, {Chu}, {Feng},
  {Gong}, {Hou}, {Hu}, {Hu}, {Hu}, {Jia}, {Jiang}, {Jiang}, {Jiang}, {Jin},
  {Li}, {Li}, {Li}, {Liu}, {Liu}, {Lu}, {Mao}, {Men}, {Qi}, {Qi}, {Shi},
  {Tang}, {Tao}, {Wang}, {Wang}, {Wang}, {Wang}, {Wang}, {Wang}, {Wang},
  {Wang}, {Wang}, {Wang}, {Wang}, {Wang}, {Xu}, {Xu}, {Yang}, {Yu}, {Yuan},
  {Yuan}, {Zhai}, {Zhang}, {Zhang}, {Zhang}, {Zhao}, {Zhou}, {Zhou}, {Zhu}, \&
  {Zou}}]{2012RAA....12.1197C}
{Cui}, X.-Q., {Zhao}, Y.-H., {Chu}, Y.-Q., {et~al.} 2012, Research in Astronomy
  and Astrophysics, 12, 1197, \dodoi{10.1088/1674-4527/12/9/003}

\bibitem[{{Eggleton} {et~al.}(1989){Eggleton}, {Fitchett}, \&
  {Tout}}]{1989ApJ...347..998E}
{Eggleton}, P.~P., {Fitchett}, M.~J., \& {Tout}, C.~A. 1989, \apj, 347, 998,
  \dodoi{10.1086/168190}

\bibitem[{{Eker} {et~al.}(2018){Eker}, {Bak{\i}{\c{s}}}, {Bilir}, {Soydugan},
  {Steer}, {Soydugan}, {Bak{\i}{\c{s}}}, {Ali{\c{c}}avu{\c{s}}}, {Aslan}, \&
  {Alpsoy}}]{2018MNRAS.479.5491E}
{Eker}, Z., {Bak{\i}{\c{s}}}, V., {Bilir}, S., {et~al.} 2018, \mnras, 479,
  5491, \dodoi{10.1093/mnras/sty1834}

\bibitem[{{El-Badry} {et~al.}(2021){El-Badry}, {Quataert}, {Rix}, {Weisz},
  {Kupfer}, {Shen}, {Xiang}, {Yang}, \& {Liu}}]{2021MNRAS.505.2051E}
{El-Badry}, K., {Quataert}, E., {Rix}, H.-W., {et~al.} 2021, \mnras, 505, 2051,
  \dodoi{10.1093/mnras/stab1318}

\bibitem[{{Gaia Collaboration} {et~al.}(2018){Gaia Collaboration}, {Brown},
  {Vallenari}, {Prusti}, {de Bruijne}, {Babusiaux}, {Bailer-Jones}, {Biermann},
  {Evans}, {Eyer}, {Jansen}, {Jordi}, {Klioner}, {Lammers}, {Lindegren},
  {Luri}, {Mignard}, {Panem}, {Pourbaix}, {Randich}, {Sartoretti}, {Siddiqui},
  {Soubiran}, {van Leeuwen}, {Walton}, {Arenou}, {Bastian}, {Cropper},
  {Drimmel}, {Katz}, {Lattanzi}, {Bakker}, {Cacciari}, {Casta{\~n}eda},
  {Chaoul}, {Cheek}, {De Angeli}, {Fabricius}, {Guerra}, {Holl}, {Masana},
  {Messineo}, {Mowlavi}, {Nienartowicz}, {Panuzzo}, {Portell}, {Riello},
  {Seabroke}, {Tanga}, {Th{\'e}venin}, {Gracia-Abril}, {Comoretto},
  {Garcia-Reinaldos}, {Teyssier}, {Altmann}, {Andrae}, {Audard},
  {Bellas-Velidis}, {Benson}, {Berthier}, {Blomme}, {Burgess}, {Busso},
  {Carry}, {Cellino}, {Clementini}, {Clotet}, {Creevey}, {Davidson}, {De
  Ridder}, {Delchambre}, {Dell'Oro}, {Ducourant},
  {Fern{\'a}ndez-Hern{\'a}ndez}, {Fouesneau}, {Fr{\'e}mat}, {Galluccio},
  {Garc{\'\i}a-Torres}, {Gonz{\'a}lez-N{\'u}{\~n}ez}, {Gonz{\'a}lez-Vidal},
  {Gosset}, {Guy}, {Halbwachs}, {Hambly}, {Harrison}, {Hern{\'a}ndez},
  {Hestroffer}, {Hodgkin}, {Hutton}, {Jasniewicz}, {Jean-Antoine-Piccolo},
  {Jordan}, {Korn}, {Krone-Martins}, {Lanzafame}, {Lebzelter}, {L{\"o}ffler},
  {Manteiga}, {Marrese}, {Mart{\'\i}n-Fleitas}, {Moitinho}, {Mora}, {Muinonen},
  {Osinde}, {Pancino}, {Pauwels}, {Petit}, {Recio-Blanco}, {Richards},
  {Rimoldini}, {Robin}, {Sarro}, {Siopis}, {Smith}, {Sozzetti}, {S{\"u}veges},
  {Torra}, {van Reeven}, {Abbas}, {Abreu Aramburu}, {Accart}, {Aerts},
  {Altavilla}, {{\'A}lvarez}, {Alvarez}, {Alves}, {Anderson}, {Andrei},
  {Anglada Varela}, {Antiche}, {Antoja}, {Arcay}, {Astraatmadja}, {Bach},
  {Baker}, {Balaguer-N{\'u}{\~n}ez}, {Balm}, {Barache}, {Barata}, {Barbato},
  {Barblan}, {Barklem}, {Barrado}, {Barros}, {Barstow}, {Bartholom{\'e}
  Mu{\~n}oz}, {Bassilana}, {Becciani}, {Bellazzini}, {Berihuete}, {Bertone},
  {Bianchi}, {Bienaym{\'e}}, {Blanco-Cuaresma}, {Boch}, {Boeche}, {Bombrun},
  {Borrachero}, {Bossini}, {Bouquillon}, {Bourda}, {Bragaglia}, {Bramante},
  {Breddels}, {Bressan}, {Brouillet}, {Br{\"u}semeister}, {Brugaletta},
  {Bucciarelli}, {Burlacu}, {Busonero}, {Butkevich}, {Buzzi}, {Caffau},
  {Cancelliere}, {Cannizzaro}, {Cantat-Gaudin}, {Carballo}, {Carlucci},
  {Carrasco}, {Casamiquela}, {Castellani}, {Castro-Ginard}, {Charlot},
  {Chemin}, {Chiavassa}, {Cocozza}, {Costigan}, {Cowell}, {Crifo}, {Crosta},
  {Crowley}, {Cuypers}, {Dafonte}, {Damerdji}, {Dapergolas}, {David}, {David},
  {de Laverny}, {De Luise}, {De March}, {de Martino}, {de Souza}, {de Torres},
  {Debosscher}, {del Pozo}, {Delbo}, {Delgado}, {Delgado}, {Di Matteo},
  {Diakite}, {Diener}, {Distefano}, {Dolding}, {Drazinos}, {Dur{\'a}n},
  {Edvardsson}, {Enke}, {Eriksson}, {Esquej}, {Eynard Bontemps}, {Fabre},
  {Fabrizio}, {Faigler}, {Falc{\~a}o}, {Farr{\`a}s Casas}, {Federici},
  {Fedorets}, {Fernique}, {Figueras}, {Filippi}, {Findeisen}, {Fonti},
  {Fraile}, {Fraser}, {Fr{\'e}zouls}, {Gai}, {Galleti}, {Garabato},
  {Garc{\'\i}a-Sedano}, {Garofalo}, {Garralda}, {Gavel}, {Gavras}, {Gerssen},
  {Geyer}, {Giacobbe}, {Gilmore}, {Girona}, {Giuffrida}, {Glass}, {Gomes},
  {Granvik}, {Gueguen}, {Guerrier}, {Guiraud}, {Guti{\'e}rrez-S{\'a}nchez},
  {Haigron}, {Hatzidimitriou}, {Hauser}, {Haywood}, {Heiter}, {Helmi}, {Heu},
  {Hilger}, {Hobbs}, {Hofmann}, {Holland}, {Huckle}, {Hypki}, {Icardi},
  {Jan{\ss}en}, {Jevardat de Fombelle}, {Jonker}, {Juh{\'a}sz}, {Julbe},
  {Karampelas}, {Kewley}, {Klar}, {Kochoska}, {Kohley}, {Kolenberg},
  {Kontizas}, {Kontizas}, {Koposov}, {Kordopatis}, {Kostrzewa-Rutkowska},
  {Koubsky}, {Lambert}, {Lanza}, {Lasne}, {Lavigne}, {Le Fustec}, {Le
  Poncin-Lafitte}, {Lebreton}, {Leccia}, {Leclerc}, {Lecoeur-Taibi},
  {Lenhardt}, {Leroux}, {Liao}, {Licata}, {Lindstr{\o}m}, {Lister}, {Livanou},
  {Lobel}, {L{\'o}pez}, {Managau}, {Mann}, {Mantelet}, {Marchal}, {Marchant},
  {Marconi}, {Marinoni}, {Marschalk{\'o}}, {Marshall}, {Martino}, {Marton},
  {Mary}, {Massari}, {Matijevi{\v{c}}}, {Mazeh}, {McMillan}, {Messina},
  {Michalik}, {Millar}, {Molina}, {Molinaro}, {Moln{\'a}r}, {Montegriffo},
  {Mor}, {Morbidelli}, {Morel}, {Morris}, {Mulone}, {Muraveva}, {Musella},
  {Nelemans}, {Nicastro}, {Noval}, {O'Mullane}, {Ord{\'e}novic},
  {Ord{\'o}{\~n}ez-Blanco}, {Osborne}, {Pagani}, {Pagano}, {Pailler},
  {Palacin}, {Palaversa}, {Panahi}, {Pawlak}, {Piersimoni}, {Pineau}, {Plachy},
  {Plum}, {Poggio}, {Poujoulet}, {Pr{\v{s}}a}, {Pulone}, {Racero}, {Ragaini},
  {Rambaux}, {Ramos-Lerate}, {Regibo}, {Reyl{\'e}}, {Riclet}, {Ripepi}, {Riva},
  {Rivard}, {Rixon}, {Roegiers}, {Roelens}, {Romero-G{\'o}mez}, {Rowell},
  {Royer}, {Ruiz-Dern}, {Sadowski}, {Sagrist{\`a} Sell{\'e}s}, {Sahlmann},
  {Salgado}, {Salguero}, {Sanna}, {Santana-Ros}, {Sarasso}, {Savietto},
  {Schultheis}, {Sciacca}, {Segol}, {Segovia}, {S{\'e}gransan}, {Shih},
  {Siltala}, {Silva}, {Smart}, {Smith}, {Solano}, {Solitro}, {Sordo}, {Soria
  Nieto}, {Souchay}, {Spagna}, {Spoto}, {Stampa}, {Steele},
  {Steidelm{\"u}ller}, {Stephenson}, {Stoev}, {Suess}, {Surdej}, {Szabados},
  {Szegedi-Elek}, {Tapiador}, {Taris}, {Tauran}, {Taylor}, {Teixeira},
  {Terrett}, {Teyssandier}, {Thuillot}, {Titarenko}, {Torra Clotet}, {Turon},
  {Ulla}, {Utrilla}, {Uzzi}, {Vaillant}, {Valentini}, {Valette}, {van Elteren},
  {Van Hemelryck}, {van Leeuwen}, {Vaschetto}, {Vecchiato}, {Veljanoski},
  {Viala}, {Vicente}, {Vogt}, {von Essen}, {Voss}, {Votruba}, {Voutsinas},
  {Walmsley}, {Weiler}, {Wertz}, {Wevers}, {Wyrzykowski}, {Yoldas},
  {{\v{Z}}erjal}, {Ziaeepour}, {Zorec}, {Zschocke}, {Zucker}, {Zurbach}, \&
  {Zwitter}}]{2018gaia}
{Gaia Collaboration}, {Brown}, A.~G.~A., {Vallenari}, A., {et~al.} 2018, \aap,
  616, A1, \dodoi{10.1051/0004-6361/201833051}

\bibitem[{{Gies} \& {Bolton}(1986)}]{1986ApJ...304..371G}
{Gies}, D.~R., \& {Bolton}, C.~T. 1986, \apj, 304, 371, \dodoi{10.1086/164171}

\bibitem[{{Green} {et~al.}(2015){Green}, {Schlafly}, {Finkbeiner}, {Rix},
  {Martin}, {Burgett}, {Draper}, {Flewelling}, {Hodapp}, {Kaiser}, {Kudritzki},
  {Magnier}, {Metcalfe}, {Price}, {Tonry}, \&
  {Wainscoat}}]{2015ApJ...810...25G}
{Green}, G.~M., {Schlafly}, E.~F., {Finkbeiner}, D.~P., {et~al.} 2015, \apj,
  810, 25, \dodoi{10.1088/0004-637X/810/1/25}

\bibitem[{{Ilijic} {et~al.}(2004){Ilijic}, {Hensberge}, {Pavlovski}, \&
  {Freyhammer}}]{2004ASPC..318..111I}
{Ilijic}, S., {Hensberge}, H., {Pavlovski}, K., \& {Freyhammer}, L.~M. 2004, in
  Astronomical Society of the Pacific Conference Series, Vol. 318,
  Spectroscopically and Spatially Resolving the Components of the Close Binary
  Stars, ed. R.~W. {Hilditch}, H.~{Hensberge}, \& K.~{Pavlovski}, 111--113

\bibitem[{{Jayasinghe} {et~al.}(2021){Jayasinghe}, {Stanek}, {Thompson},
  {Kochanek}, {Rowan}, {Vallely}, {Strassmeier}, {Weber}, {Hinkle}, {Hambsch},
  {Martin}, {Prieto}, {Pessi}, {Huber}, {Auchettl}, {Lopez}, {Ilyin},
  {Badenes}, {Howard}, {Isaacson}, \& {Murphy}}]{2021arXiv210102212J}
{Jayasinghe}, T., {Stanek}, K.~Z., {Thompson}, T.~A., {et~al.} 2021, \mnras,
  504, 2577, \dodoi{10.1093/mnras/stab907}

\bibitem[{{Kiziltan} {et~al.}(2013){Kiziltan}, {Kottas}, {De Yoreo}, \&
  {Thorsett}}]{2013ApJ...778...66K}
{Kiziltan}, B., {Kottas}, A., {De Yoreo}, M., \& {Thorsett}, S.~E. 2013, \apj,
  778, 66, \dodoi{10.1088/0004-637X/778/1/66}

\bibitem[{{Kovalev} {et~al.}(2022){Kovalev}, {Xuefei}, \&
  {Zhanwen}}]{2022arXiv220706996K}
{Kovalev}, M., {Xuefei}, C., \& {Zhanwen}, H. 2022, arXiv e-prints,
  arXiv:2207.06996.
\newblock \doarXiv{2207.06996}

\bibitem[{{Kurucz}(1993)}]{Kurucz1993}
{Kurucz}, R.~L. 1993, {SYNTHE spectrum synthesis programs and line data}

\bibitem[{{Li} {et~al.}(2022){Li}, {Wang}, {Zhao}, {Bai}, {Yuan}, {Zhang}, \&
  {Liu}}]{2022arXiv220900765X}
{Li}, X., {Wang}, S., {Zhao}, X., {et~al.} 2022, arXiv e-prints,
  arXiv:2209.00765.
\newblock \doarXiv{2209.00765}

\bibitem[{{Linares} {et~al.}(2018){Linares}, {Shahbaz}, \&
  {Casares}}]{2018ApJ...859...54L}
{Linares}, M., {Shahbaz}, T., \& {Casares}, J. 2018, \apj, 859, 54,
  \dodoi{10.3847/1538-4357/aabde6}

\bibitem[{{Liu} {et~al.}(2020){Liu}, {Fu}, {Shi}, {Wu}, {Han}, {Chen}, {Dong},
  {Zhao}, {Chen}, {Zhang}, {Bai}, {Chen}, {Cui}, {Du}, {Hsia}, {Jiang}, {Hou},
  {Hou}, {Li}, {Li}, {Li}, {Liu}, {Liu}, {Luo}, {Ren}, {Tian}, {Tian}, {Wang},
  {Wu}, {Xie}, {Yan}, {Yang}, {Yu}, {Zhang}, {Zhang}, {Zhang}, {Zhang}, {Zhao},
  {Zhong}, {Zong}, \& {Zuo}}]{2020arXiv200507210L}
{Liu}, C., {Fu}, J., {Shi}, J., {et~al.} 2020, arXiv e-prints,
  arXiv:2005.07210.
\newblock \doarXiv{2005.07210}

\bibitem[{{Liu} {et~al.}(2019){Liu}, {Zhang}, {Howard}, {Bai}, {Lu}, {Soria},
  {Justham}, {Li}, {Zheng}, {Wang}, {Belczynski}, {Casares}, {Zhang}, {Yuan},
  {Dong}, {Lei}, {Isaacson}, {Wang}, {Bai}, {Shao}, {Gao}, {Wang}, {Niu},
  {Cui}, {Zheng}, {Mu}, {Zhang}, {Wang}, {Heger}, {Qi}, {Liao}, {Lattanzi},
  {Gu}, {Wang}, {Wu}, {Shao}, {Shen}, {Wang}, {Bregman}, {Di Stefano}, {Liu},
  {Han}, {Zhang}, {Wang}, {Ren}, {Zhang}, {Zhang}, {Wang}, {Cabrera-Lavers},
  {Corradi}, {Rebolo}, {Zhao}, {Zhao}, {Chu}, \& {Cui}}]{2019Natur.575..618L}
{Liu}, J., {Zhang}, H., {Howard}, A.~W., {et~al.} 2019, \nat, 575, 618,
  \dodoi{10.1038/s41586-019-1766-2}

\bibitem[{{Luo} {et~al.}(2015){Luo}, {Zhao}, {Zhao}, {Deng}, {Liu}, {Jing},
  {Wang}, {Zhang}, {Shi}, {Cui}, {Chu}, {Li}, {Bai}, {Wu}, {Cai}, {Cao}, {Cao},
  {Carlin}, {Chen}, {Chen}, {Chen}, {Chen}, {Chen}, {Chen}, {Chen},
  {Christlieb}, {Chu}, {Cui}, {Dong}, {Du}, {Fan}, {Feng}, {Fu}, {Gao}, {Gong},
  {Gu}, {Guo}, {Han}, {He}, {Hou}, {Hou}, {Hou}, {Hu}, {Hu}, {Hu}, {Huo},
  {Jia}, {Jiang}, {Jiang}, {Jiang}, {Jin}, {Kong}, {Kong}, {Lei}, {Li}, {Li},
  {Li}, {Li}, {Li}, {Li}, {Li}, {Li}, {Li}, {Li}, {Li}, {Li}, {Liang}, {Lin},
  {Liu}, {Liu}, {Liu}, {Liu}, {Lu}, {Luo}, {Mao}, {Newberg}, {Ni}, {Qi}, {Qi},
  {Shen}, {Shi}, {Song}, {Song}, {Su}, {Su}, {Tang}, {Tao}, {Tian}, {Wang},
  {Wang}, {Wang}, {Wang}, {Wang}, {Wang}, {Wang}, {Wang}, {Wang}, {Wang},
  {Wang}, {Wang}, {Wang}, {Wang}, {Wang}, {Wang}, {Wang}, {Wang}, {Wang},
  {Wang}, {Wei}, {Wei}, {Wu}, {Wu}, {Wu}, {Wu}, {Xing}, {Xu}, {Xu}, {Xu},
  {Yan}, {Yang}, {Yang}, {Yang}, {Yang}, {Yao}, {Yu}, {Yuan}, {Yuan}, {Yuan},
  {Yuan}, {Zhai}, {Zhang}, {Zhang}, {Zhang}, {Zhang}, {Zhang}, {Zhang},
  {Zhang}, {Zhang}, {Zhao}, {Zhou}, {Zhou}, {Zhu}, {Zhu}, {Zou}, \&
  {Zuo}}]{2015RAA....15.1095L}
{Luo}, A.~L., {Zhao}, Y.-H., {Zhao}, G., {et~al.} 2015, Research in Astronomy
  and Astrophysics, 15, 1095, \dodoi{10.1088/1674-4527/15/8/002}

\bibitem[{{Magic} {et~al.}(2010){Magic}, {Serenelli}, {Weiss}, \&
  {Chaboyer}}]{2010ApJ...718.1378M}
{Magic}, Z., {Serenelli}, A., {Weiss}, A., \& {Chaboyer}, B. 2010, \apj, 718,
  1378, \dodoi{10.1088/0004-637X/718/2/1378}

\bibitem[{{Margalit} \& {Metzger}(2017)}]{2017ApJ...850L..19M}
{Margalit}, B., \& {Metzger}, B.~D. 2017, \apjl, 850, L19,
  \dodoi{10.3847/2041-8213/aa991c}

\bibitem[{{Mazeh} {et~al.}(2022){Mazeh}, {Faigler}, {Bashi}, {Shahaf},
  {Davidson}, {Green}, {Gomel}, {Maoz}, {Sussholz}, {Dong}, {Zhang}, {Liu},
  {Wang}, {Luo}, {Zheng}, {Hallakoun}, {Perdelwitz}, {Latham}, {Ribas},
  {Baroch}, {Morales}, {Nagel}, {Santos}, {Ciardi}, {Christiansen}, {Lund}, \&
  {Winn}}]{2022arXiv220611270M}
{Mazeh}, T., {Faigler}, S., {Bashi}, D., {et~al.} 2022, arXiv e-prints,
  arXiv:2206.11270.
\newblock \doarXiv{2206.11270}

\bibitem[{{McGale} {et~al.}(1995){McGale}, {Allan}, {Barber}, {Bertram},
  {Denby}, {Page}, {Pye}, {Ricketts}, {Stewart}, {Watson}, \&
  {West}}]{1995AdSpR..16c..69M}
{McGale}, P.~A., {Allan}, D.~J., {Barber}, C.~R., {et~al.} 1995, Advances in
  Space Research, 16, 69

\bibitem[{{Monet} {et~al.}(2003){Monet}, {Levine}, {Canzian}, {Ables}, {Bird},
  {Dahn}, {Guetter}, {Harris}, {Henden}, {Leggett}, {Levison}, {Luginbuhl},
  {Martini}, {Monet}, {Munn}, {Pier}, {Rhodes}, {Riepe}, {Sell}, {Stone},
  {Vrba}, {Walker}, {Westerhout}, {Brucato}, {Reid}, {Schoening}, {Hartley},
  {Read}, \& {Tritton}}]{2003AJ....125..984M}
{Monet}, D.~G., {Levine}, S.~E., {Canzian}, B., {et~al.} 2003, \aj, 125, 984,
  \dodoi{10.1086/345888}

\bibitem[{{Montet} {et~al.}(2015){Montet}, {Morton}, {Foreman-Mackey},
  {Johnson}, {Hogg}, {Bowler}, {Latham}, {Bieryla}, \&
  {Mann}}]{2015ApJ...809...25M}
{Montet}, B.~T., {Morton}, T.~D., {Foreman-Mackey}, D., {et~al.} 2015, \apj,
  809, 25, \dodoi{10.1088/0004-637X/809/1/25}

\bibitem[{{Morton}(2015)}]{2015ascl.soft03010M}
{Morton}, T.~D. 2015, {isochrones: Stellar model grid package}.
\newblock \doeprint{1503.010}

\bibitem[{{Orosz} \& {Hauschildt}(2000)}]{2000A&A...364..265O}
{Orosz}, J.~A., \& {Hauschildt}, P.~H. 2000, \aap, 364, 265.
\newblock \doarXiv{astro-ph/0010114}

\bibitem[{{Pooley} {et~al.}(2018){Pooley}, {Kumar}, {Wheeler}, \&
  {Grossan}}]{2018ApJ...859L..23P}
{Pooley}, D., {Kumar}, P., {Wheeler}, J.~C., \& {Grossan}, B. 2018, \apjl, 859,
  L23, \dodoi{10.3847/2041-8213/aac3d6}

\bibitem[{{Price-Whelan} {et~al.}(2017){Price-Whelan}, {Hogg},
  {Foreman-Mackey}, \& {Rix}}]{2017ApJ...837...20P}
{Price-Whelan}, A.~M., {Hogg}, D.~W., {Foreman-Mackey}, D., \& {Rix}, H.-W.
  2017, \apj, 837, 20, \dodoi{10.3847/1538-4357/aa5e50}

\bibitem[{{Remillard} \& {McClintock}(2006)}]{2006ARA&A..44...49R}
{Remillard}, R.~A., \& {McClintock}, J.~E. 2006, \araa, 44, 49,
  \dodoi{10.1146/annurev.astro.44.051905.092532}

\bibitem[{{Rivinius} {et~al.}(2020){Rivinius}, {Baade}, {Hadrava}, {Heida}, \&
  {Klement}}]{2020A&A...637L...3R}
{Rivinius}, T., {Baade}, D., {Hadrava}, P., {Heida}, M., \& {Klement}, R. 2020,
  \aap, 637, L3, \dodoi{10.1051/0004-6361/202038020}

\bibitem[{{Shao} \& {Li}(2019)}]{2019ApJ...885..151S}
{Shao}, Y., \& {Li}, X.-D. 2019, \apj, 885, 151,
  \dodoi{10.3847/1538-4357/ab4816}

\bibitem[{{Shibata} {et~al.}(2019){Shibata}, {Zhou}, {Kiuchi}, \&
  {Fujibayashi}}]{2019PhRvD.100b3015S}
{Shibata}, M., {Zhou}, E., {Kiuchi}, K., \& {Fujibayashi}, S. 2019, \prd, 100,
  023015, \dodoi{10.1103/PhysRevD.100.023015}

\bibitem[{{Sneden}(1973)}]{Sneden1973}
{Sneden}, C.~A. 1973, PhD thesis, THE UNIVERSITY OF TEXAS AT AUSTIN.

\bibitem[{{Thompson} {et~al.}(2019){Thompson}, {Kochanek}, {Stanek}, {Badenes},
  {Post}, {Jayasinghe}, {Latham}, {Bieryla}, {Esquerdo}, {Berlind}, {Calkins},
  {Tayar}, {Lindegren}, {Johnson}, {Holoien}, {Auchettl}, \&
  {Covey}}]{2019Sci...366..637T}
{Thompson}, T.~A., {Kochanek}, C.~S., {Stanek}, K.~Z., {et~al.} 2019, Science,
  366, 637, \dodoi{10.1126/science.aau4005}

\bibitem[{{van Kerkwijk} {et~al.}(2011){van Kerkwijk}, {Breton}, \&
  {Kulkarni}}]{2011ApJ...728...95V}
{van Kerkwijk}, M.~H., {Breton}, R.~P., \& {Kulkarni}, S.~R. 2011, \apj, 728,
  95, \dodoi{10.1088/0004-637X/728/2/95}

\bibitem[{{Wade} \& {Horne}(1988)}]{1988ApJ...324..411W}
{Wade}, R.~A., \& {Horne}, K. 1988, \apj, 324, 411, \dodoi{10.1086/165905}

\bibitem[{{Wang} {et~al.}(2021){Wang}, {Zhang}, {Bai}, {Yuan}, {Xiang},
  {Zhang}, {Hou}, {Zuo}, {Du}, {Li}, {Yang}, {Cui}, {Wang}, {Li}, {Kovalev},
  {Li}, {Tian}, {Zong}, {Han}, {Liu}, {Luo}, {Shi}, {Fu}, {Bi}, {Han}, \&
  {Liu}}]{2021RAA....21..292W}
{Wang}, S., {Zhang}, H.-T., {Bai}, Z.-R., {et~al.} 2021, Research in Astronomy
  and Astrophysics, 21, 292, \dodoi{10.1088/1674-4527/21/11/292}

\bibitem[{{Wilson} \& {Devinney}(1971)}]{1971ApJ...166..605W}
{Wilson}, R.~E., \& {Devinney}, E.~J. 1971, \apj, 166, 605,
  \dodoi{10.1086/150986}

\bibitem[{{Yan} {et~al.}(2022){Yan}, {Li}, {Wang}, {Zong}, {Yuan}, {Xiang},
  {Huang}, {Xie}, {Dong}, {Yuan}, {Bi}, {Chu}, {Cui}, {Deng}, {Fu}, {Han},
  {Hou}, {Li}, {Liu}, {Liu}, {Liu}, {Luo}, {Shi}, {Wu}, {Zhang}, {Zhao}, \&
  {Zhao}}]{2022Innov...300224Y}
{Yan}, H., {Li}, H., {Wang}, S., {et~al.} 2022, The Innovation, 3, 100224,
  \dodoi{10.1016/j.xinn.2022.100224}

\bibitem[{{Yi} {et~al.}(2019){Yi}, {Sun}, \& {Gu}}]{2019ApJ...886...97Y}
{Yi}, T., {Sun}, M., \& {Gu}, W.-M. 2019, \apj, 886, 97,
  \dodoi{10.3847/1538-4357/ab4a75}

\bibitem[{{Yi} {et~al.}(2022){Yi}, {Gu}, {Zhang}, {Zheng}, {Sun}, {Wang},
  {Bai}, {Wang}, {Wu}, {Bai}, {Wang}, {Zhang}, {Dong}, {Shao}, {Li}, {Zhang},
  {Huang}, {Yang}, {Yu}, {Mu}, {Fu}, {Qi}, {Guo}, {Fang}, {Zheng}, {Li}, {Shi},
  {Chen}, \& {Liu}}]{2022NatAs.tmp..201Y}
{Yi}, T., {Gu}, W.-M., {Zhang}, Z.-X., {et~al.} 2022, Nature Astronomy,
  \dodoi{10.1038/s41550-022-01766-0}

\bibitem[{{Yu} {et~al.}(2018){Yu}, {Liu}, \& {Dai}}]{2018ApJ...861..114Y}
{Yu}, Y.-W., {Liu}, L.-D., \& {Dai}, Z.-G. 2018, \apj, 861, 114,
  \dodoi{10.3847/1538-4357/aac6e5}

\bibitem[{{Zacharias} {et~al.}(2012){Zacharias}, {Finch}, {Girard}, {Henden},
  {Bartlett}, {Monet}, \& {Zacharias}}]{2012yCat.1322....0Z}
{Zacharias}, N., {Finch}, C.~T., {Girard}, T.~M., {et~al.} 2012, VizieR Online
  Data Catalog, I/322A

\bibitem[{{Zacharias} {et~al.}(2013){Zacharias}, {Finch}, {Girard}, {Henden},
  {Bartlett}, {Monet}, \& {Zacharias}}]{2013AJ....145...44Z}
---. 2013, \aj, 145, 44, \dodoi{10.1088/0004-6256/145/2/44}

\bibitem[{{Zhao} {et~al.}(2012){Zhao}, {Zhao}, {Chu}, {Jing}, \&
  {Deng}}]{2012RAA....12..723Z}
{Zhao}, G., {Zhao}, Y.-H., {Chu}, Y.-Q., {Jing}, Y.-P., \& {Deng}, L.-C. 2012,
  Research in Astronomy and Astrophysics, 12, 723,
  \dodoi{10.1088/1674-4527/12/7/002}

\bibitem[{{Zhou} {et~al.}(2009){Zhou}, {Jiang}, {Zhang}, \&
  {Wei}}]{2009RAA.....9..349Z}
{Zhou}, A.-Y., {Jiang}, X.-J., {Zhang}, Y.-P., \& {Wei}, J.-Y. 2009, Research
  in Astronomy and Astrophysics, 9, 349, \dodoi{10.1088/1674-4527/9/3/010}

\bibitem[{{Zong} {et~al.}(2020){Zong}, {Fu}, {De Cat}, {Wang}, {Shi}, {Luo},
  {Zhang}, {Frasca}, {Molenda-{\.Z}akowicz}, {Gray}, {Corbally}, {Catanzaro},
  {Cang}, {Wang}, {Chen}, {Hou}, {Liu}, {Niu}, {Pan}, {Tian}, {Yan}, {Zhang},
  \& {Zuo}}]{2020ApJS..251...15Z}
{Zong}, W., {Fu}, J.-N., {De Cat}, P., {et~al.} 2020, \apjs, 251, 15,
  \dodoi{10.3847/1538-4365/abbb2d}

\end{thebibliography}

\end{document}